\documentclass[%
 reprint,superscriptaddress,
nobibnotes,
 amsmath,amssymb,
 aps,
 prb,
]{revtex4-2}
\pdfoutput=1
\usepackage[ngerman, english]{babel}
\usepackage{bm}%
\usepackage[dvipsnames]{xcolor}
\usepackage{graphicx}
\usepackage[utf8]{inputenc}
\usepackage{epstopdf}
\usepackage{afterpage}
\usepackage{setspace}
\usepackage{booktabs}
\usepackage{hyperref}
\usepackage{placeins}
\usepackage{tabularx}

\expandafter\ifx\csname package@font\endcsname\relax\else
 \expandafter\expandafter
 \expandafter\usepackage
 \expandafter\expandafter
 \expandafter{\csname package@font\endcsname}%
\fi
\begin{document}

\setstretch{1.0}

\title{Free energy profiles for chemical reactions in solution from high-dimensional neural network potentials: The case of the Strecker synthesis}

\author{Alea Miako Tokita}
\email{alea.tokita@ruhr-uni-bochum.de}
\affiliation{Lehrstuhl f\"ur Theoretische Chemie II, Ruhr-Universit\"at Bochum, 44780 Bochum, Germany}
\affiliation{Research Center Chemical Sciences and Sustainability, Research Alliance Ruhr, 44780 Bochum, Germany}
\author{Timothée Devergne}
\affiliation{Institut de Minéralogie, de Physique des Matériaux et de Cosmochimie, Sorbonne Université, Paris, France}
\affiliation{Istituto Italiano di Tecnologia, Genova, Italy}
\author{A. Marco Saitta}
\affiliation{Institut de Minéralogie, de Physique des Matériaux et de Cosmochimie, Sorbonne Université, Paris, France}
\author{Jörg Behler}
\email{joerg.behler@ruhr-uni-bochum.de}
\affiliation{Lehrstuhl f\"ur Theoretische Chemie II, Ruhr-Universit\"at Bochum, 44780 Bochum, Germany}
\affiliation{Research Center Chemical Sciences and Sustainability, Research Alliance Ruhr, 44780 Bochum, Germany}

\date{\today}

\begin{abstract}
Machine learning potentials (MLPs) have become a popular tool in chemistry and materials science as they combine the accuracy of electronic structure calculations with the high computational efficiency of analytic potentials. 
MLPs are particularly useful for computationally demanding simulations such as the determination of free energy profiles governing chemical reactions in solution, but to date such applications are still rare. In this work we show how umbrella sampling simulations can be combined with active learning of high-dimensional neural network potentials (HDNNPs) to construct free energy profiles in a systematic way. For the example of the first step of Strecker synthesis of glycine in aqueous solution we provide a detailed analysis of the improving quality of HDNNPs for datasets of increasing size. 
We find that next to the typical quantification of energy and force errors with respect to the underlying density functional theory data also the long-term stability of the simulations and the convergence of physical properties should be rigorously monitored to obtain reliable and converged free energy profiles of chemical reactions in solution.

\end{abstract}

\maketitle 

\section{Introduction}\label{sec:introduction}

Reactions in solution are of fundamental importance in chemistry, ranging from the synthesis of small organic molecules and pharmaceuticals to complex biomolecular processes. The solvent molecules play a crucial role by influencing reaction rates and yields, chemical equilibria, and product selectivities~\cite{rosskyDynamicsChemicalProcesses1994, reichardtSolventsSolventEffects2011, orr-ewingTakingPlungeChemical2017, patrickSolventEffectsAnion2024}. 
Accurately describing such reactions in computer simulations necessitates the use of quantum mechanical methods, such as density functional theory (DFT)~\cite{P3051, burkeDFTNutshell2013}. However, theoretical studies of chemical processes in solution using DFT are computationally demanding due to the large number of solvent molecules required to realistically model the molecular solvation environment.
Additionally, the liquid solvent is a dynamic system that must be adequately sampled at finite temperatures to obtain free energy profiles governing the reactions. 
Enhanced sampling techniques~\cite{laio_escaping_2002, invernizzi_rethinking_2020, torrie_nonphysical_1977}, which add a bias to the potential energy, can be used to study chemical reactions with reduced computational costs, but still these calculations remain very demanding. 
Consequently, using electronic structure methods in \textit{ab initio} molecular dynamics (AIMD) simulations~\cite{P0433,B0007} directly is very challenging and feasible only for very simple systems on short timescales.\\ 

Nowadays, machine learning potentials (MLPs)~\cite{P4885,P6121,P6102,P6112,P5793,P4263,P5673,P5788,P4444,P6131,P3033,P2559,P5977,P5877,yangMachineLearningReactive2024} have emerged as a tool to retain the high accuracy of electronic structure methods at strongly reduced computational costs. A wide range of methods is available, including neural network potentials~\cite{P1174,smithANI1ExtensibleNeural2017b,wangDeePMDkitDeepLearning2018}, kernel-based approaches~\cite{bartokGaussianApproximationPotentials2010b,chmielaMachineLearningAccurate2017, chmielaExactMolecularDynamics2018}, atomic cluster expansion~\cite{drautzAtomicClusterExpansion2019b, kovacsLinearAtomicCluster2021} and message-passing neural networks~\cite{pmlr-v70-gilmer17a, schuttSchNetDeepLearning2018,unkePhysNetNeuralNetwork2019a,batznerE3equivariantGraphNeural2022,NEURIPS2022_4a36c3c5}.
MLPs have been successfully applied to a broad range of aqueous systems~\cite{omranpourPerspectiveAtomisticSimulations2024} including water, solvated ions and solid-liquid interfaces~\cite{P4556,P6151,daru_coupled_2022,hellstromStructureAqueousNaOH2016, xu_molecular_2019, hellstromNuclearQuantumEffects2018a,  quarantaStructureDynamicsLiquid2019, kapil_inexpensive_2020, P6271,eckhoffInsightsLithiumManganese2021a, oneill_crumbling_2022, wang_structures_2023, nakanishi_theoretical_2023, zeng_mechanistic_2023}. 
Moreover, MLPs promise great potential to study molecular reactions in solution. They have been applied in different ways such as the simulation of molecules and chemical reactions in combination with implicit solvent models\;\cite{angActiveLearningAccelerates2021, katzberger_general_2024,chen_machine_2021, noe_machine_2020} and solvents described by classical force fields\;\cite{brickelReactiveMolecularDynamics2019, shen_multiscale_2016, topfer_double_2022, zhou_accelerated_2023}. Further, also some studies of chemical reactions in explicit solvents described by MLPs at the \emph{ab initio} level have been reported\;\cite{P6573, zhangIntramolecularWaterMediated2024, zhang_modeling_2023, yang_using_2022, yangRoleWaterReaction2019,benayadPrebioticChemicalReactivity2024,huet_new_2024, Devergne2024, vitartasActiveLearningMeets2024, celerseCapturingDichotomicSolvent2024, youngTransferableActivelearningStrategy2021, anmolUnveilingRoleSolvent2024}. 

\begin{figure*}
    \centering
    \includegraphics[width=2\columnwidth]{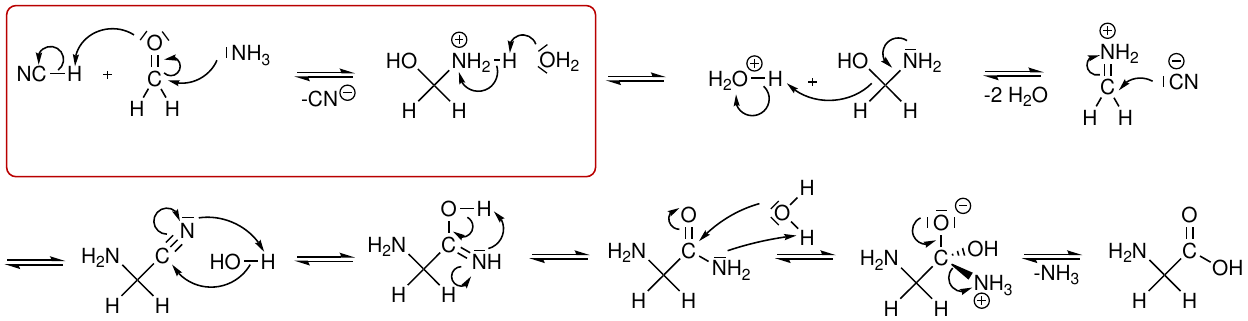}
    \caption{Mechanism of the Strecker synthesis of glycine as studied in the work of Magrino et al.\;\cite{magrino_step_2021}. In the present work we focus on the first reaction step  highlighted by the red box.} 
    \label{fig:full_strecker}
\end{figure*}

An interesting example for chemical reactions in solution is the Strecker synthesis~\cite{strecker_ueber_1854, wang_asymmetric_2011, groger_catalytic_2003} of $\alpha$-amino acids via condensation of aldehydes, amines or ammonia, and cyanides~\cite{kobayashiCatalyticEnantioselectiveAddition1999}. A  lot of effort has been put in unravelling its reaction mechanism\;\cite{ogataMechanisticAspectsStrecker1971a, tailladesSystemesStreckerApparentes1974, van1975strecker, millerStreckerSynthesisPrimitive1981, moutouEquilibriumAaminoacetonitrileFormation1995, xuTheoreticalStudiesAminoacetonitrile2007, yamabeProtonTransfersStrecker2014,thrushReactionsGlycolonitrileAmmonia2018, magrino_step_2021, chimiakIsotopeEffectsOrigin2022}.
One of these studies is the work of Devergne et. al.~\cite{P6573}, which focuses on the reaction of formaldehyde, hydrogen cyanide and ammonia to glycine in water (cf. Fig.~\ref{fig:full_strecker}). 
By making use of an MLP trained on data obtained from extensive AIMD trajectories along the reaction path~\cite{magrino_step_2021} the computational time necessary to sample the reaction free energy was strongly decreased. 
However, using costly AIMD trajectories along the full reaction path as basis for training MLPs is not very efficient since the configurations visited along the trajectories are strongly correlated. Moreover, MLP-driven MD simulations were found to be reliable only for  configurations similar to those included in the reference dataset, i.e., in the available AIMD trajectories~\cite{P6573}. Hence the accessible length of the MLP-based simulations of Strecker synthesis was restricted to the order of magnitude of the underlying AIMD simulations due to the lack of stability of the MLP caused by the limited coverage of configuration space. To increase the stability of these simulations, mirror reflection operations were required to avoid leaving the known configuration space~\cite{P6573}.
Consequently, while for systems such as pure liquid water, MLPs trained to data based on extensive AIMD simulations have been proven to enable long-term stable simulations\;\cite{stolte_random_2024}, this does not necessarily seem to be the case for more complex systems, in particular if substantial energy barriers are involved in the making and breaking of bonds. 

This limitation can be overcome by constructing the reference dataset through active learning (AL)~\cite{seungQueryCommittee1992a}. AL is a commonly used tool to explore the configuration space and construct MLPs based on only those configurations which provide new information about the potential energy surface (PES). This allows to perform demanding electronic structure calculations only for missing parts of the PES\;\cite{P6543,P3114,smith_less_2018,P5782,schran_automated_2020, schran_committee_2020}. AL is commonly used in combination with equilibrium molecular dynamics (MD). On the other hand, advanced simulation techniques such as metadynamics~\cite{laio_escaping_2002}, umbrella sampling~\cite{torrie_nonphysical_1977} or transition path sampling~\cite{bolhuis2002} are necessary to access rare events such as reaction barrier crossings. These methods enable to efficiently generate configurations needed for the construction of MLPs\;\cite{Devergne2024} and they have been used in some studies to enhance AL, mainly in combination with metadynamics\;\cite{yang_using_2022,yang_neural_2023, guan_using_2023,vitartasActiveLearningMeets2024,celerseCapturingDichotomicSolvent2024} or on-the-fly probability-enhanced sampling (OPES)\;\cite{zhangIntramolecularWaterMediated2024, david_arcann_2024, perego_data-efficient_2024}.

Using the Strecker synthesis of glycine in aqueous solution as an example, in this work we present a blueprint for the systematic construction of a high-dimensional neural network potential (HDNNP) \cite{P1174} applicable to chemical reactions in an explicit solvent involving high free energy barriers. For this purpose, AL is combined with umbrella sampling simulations\;\cite{torrie_nonphysical_1977} driven by preliminary potentials for efficient sampling of configurations along the reaction path. A particular focus of our work is on the evolution of the quality of the HDNNP in the course of AL, which is not only monitored by determining the errors of energies and forces with respect to the underlying reference DFT method, but also based on the stability in simulations, on the representation of physical properties like radial distribution functions, the coverage of configuration space, and the uncertainty in the prediction of new geometries encountered in the simulations.
The final potential, which allows to perform long-term stable simulations at low computational costs, is shown to provide a converged free energy profile of the reaction.

\section{Methods}\label{sec:methods}
\subsection{Path Collective Variables}

Sampling rare events such as crossing free energy barriers of chemical reactions often requires prohibitively long MD simulations.
The computational effort to sample the reaction path can be reduced significantly by using advanced simulation techniques, such as metadynamics\;\cite{laio_escaping_2002, barducci_well-tempered_2008} or umbrella sampling\;\cite{torrie_nonphysical_1977}. These techniques typically require the definition of collective variables (CVs), which project the full-dimensional configuration space into, e.g., a one or two-dimensional space that characterizes the progress of the reaction. Various options exist to define CVs, one of them is to make use of atomic coordination numbers\;\cite{sprik_coordination_1998}. In this work we follow the definition of Ref.~[\citenum{pietrucciFormamideReactionNetwork2015}],
in which a coordination number $C_{i}^{\alpha \sigma}$ of a central atom $i$ of element $\alpha$ depends on all atoms $j$ of element $\sigma$ as
\begin{align}
   C_{i}^{\alpha \sigma} =\sum_{j \in \sigma } \frac{1-\left[\frac{r_{i j}}{r_0^{\alpha \sigma}}\right]^8}{1-\left[\frac{r_{i j}}{r_0^{\alpha \sigma}}\right]^{14}} ~.
\end{align}
The terms in the sum decay from one to zero with increasing distance $r_{ij}$ between central atom $i$ and neighbor $j$. The onset and slope of the decay are defined by the element pair-specific parameter $r_0^{\alpha \sigma}$.
While these continuous coordination numbers in principle can be calculated for all atoms in the system, the coordination numbers of the in total $N_\mathrm{central} = 5$ carbon, nitrogen and oxygen atoms of the reactants in the first step of the Strecker synthesis (cf. Fig.~\ref{fig:full_strecker}) are most relevant for the reaction and thus selected for the computation of the CVs. As for each of these central atoms the coordination numbers are defined with respect to all $N_\mathrm{elements} = 4$ elements in the system, in total 20 coordination numbers are obtained.

In the next step, these coordination numbers are used to define a similarity measure $D$ between a structure of the system $x(t)$ at time $t$ and a reference structure $X$. $D$ is constructed as the squared difference of all coordination numbers,
\begin{align}
D\left[x(t), X\right]=\sum_{i=1}^{N_{\mathrm{central}}}\sum^{N_{\mathrm{elements}}}_{\sigma=1} \left(C_{i}^{\alpha \sigma}\left(x(t)\right)-C_{i}^{\alpha \sigma}\left(X \right)\right)^2 ~. \label{eq:D} 
\end{align}
Finally, this similarity measure is used to calculate the path CVs $s$ and $z$, which define the position of the system with respect to the reaction path based on $P$ reference structures as proposed by Branduardi et al.\;\cite{Branduardi2007}, 
\begin{align}
    s(t)=\frac{1}{P - 1}\left(\frac{\sum_{\beta=1}^P \beta \exp \left(-\lambda D\left[x(t), X_\beta\right]\right)}{\sum_{\beta=1}^P \exp \left(-\lambda D\left[x(t), X_\beta\right]\right)}-1\right)
    \label{eq:s_calc}
\end{align}
and
\begin{align}
    z(t)=-\frac{1}{\lambda} \log \sum_{\beta=1}^P \exp \left(-\lambda D\left[x(t), X_\beta\right]\right).
\end{align}
The CV $s$ describes at which point along the reaction path a configuration $x(t)$ is located. Compared to the original version of Eq.~\;\ref{eq:s_calc} in Ref.\cite{Branduardi2007}, here a scaling factor as introduced in Ref.~[\citenum{magrino_step_2021}] is included such that $s \in [0,1]$. The second CV 
$z$ measures the deviation of the configuration at time $t$ from the reaction path. Since according to Eq.~\ref{eq:D} $D$ is always positive, the exponential function $\exp{(-\lambda D)}$ monotonously decreases to zero with increasing difference in coordination numbers. The parameter $ \lambda $ is estimated from two subsequent points along the reaction path using the relation $ \exp \left(-\lambda D\left[X_\beta, X_{\beta+1}\right]\right) \approx 0.1 $ to achieve a smooth free energy landscape.

\subsection{Umbrella sampling and free energy calculation}

The free energy profile $A$ is computed with the distribution function $\langle\rho(s)\rangle$ at a position $s$ along the reaction path using
\begin{align}
    A(s)=-k_{\mathrm{B}} T \ln \langle\rho(s)\rangle\;,
    \label{eq:free_energy}
\end{align}
where $k_{\mathrm{B}}$ is the Boltzmann constant and $T$ the temperature.
In this work we use umbrella sampling as described in detail in Refs.~[\citenum{magrino_step_2021,P6573}] to efficiently access the distribution of $s$. In umbrella sampling the system is confined to $W$ umbrella sampling windows centered at specific values $s_j$ along the reaction path. By applying a quadratic bias potential acting on the $s$ path CV with strength $k$,
    \begin{align}
        V_{\text {bias}, j}(s)=\frac{k}{2}\left(s-s_j\right)^2\;,
        \label{eq:V_bias}
    \end{align}
the system is confined to its respective umbrella sampling window.

The spacing between two windows depends on $k$ as
\begin{equation}
    s_{j+1}-s_j = \sqrt{\frac{k_\mathrm{B}T}{k}}\quad ,
\end{equation}
which ensures sufficient overlap between the windows for continuous sampling along the reaction path. To constrain the system to the reaction path of interest an additional bias is applied if $z$ exceeds a predefined threshold value.

To obtain the distribution function $\langle\rho(s)\rangle$, first in each umbrella sampling window an MD simulation is carried out resulting in $W$ biased distributions of $s$. Then, the bias introduced by umbrella sampling needs to be removed for the free energy calculation. For this purpose, reweighing techniques such as the weighted histogram analysis method (WHAM) are employed~\cite{kumarMultidimensionalFreeEnergy1995}.

\subsection{High-dimensional neural network potentials}

In this work, we employ second-generation HDNNPs~\cite{P1174, P6018} to compute the energies and forces required to propagate the MD simulations. The total energy $E_{\mathrm{tot}}$ of the system is given by
\begin{align}
E_{\mathrm{tot}}=\sum_{i=1}^{N_\mathrm{atoms}} E_{i} 
\end{align}
as a sum of atomic energies $E_i$.
The atomic energies depend on the local atomic environments defined by a cutoff radius. The positions of the neighboring atoms inside this environment are described by vectors of atom-centered  symmetry functions (ACSFs)~\cite{P2882}, which fulfill the mandatory translational, rotational and permutational invariances of the energy with respect to the atomic positions. Two types of ACSFs are used, which are radial symmetry functions and angular symmetry functions. The radial symmetry functions provide a radial coordination fingerprint with respect to each element in the system while angular symmetry functions include additional angular information.  In a system containing four elements, as in this work, typically around 100 ACSF values are used in each of the vectors representing the atomic environments. These vectors then serve as input for atomic feed-forward neural networks (NN) yielding the atomic energies. For a given element, the architecture and weights of the atomic NNs are constrained to be the same to ensure that the potential energy is invariant with respect to permutation of atoms of the same element. The weight parameters are determined iteratively using total energies and atomic forces from reference DFT calculations. Further details about HDNNPs, their properties and the training process can be found in several reviews~\cite{P6018,P5128,tokitaHowTrainNeural2023,P4444}.

\section{Computational Details}\label{sec:computational}

\subsection{Construction of the reference dataset}
\label{subsec:dataset_construction}

\begin{figure}
    \centering
    \includegraphics[width=\columnwidth]{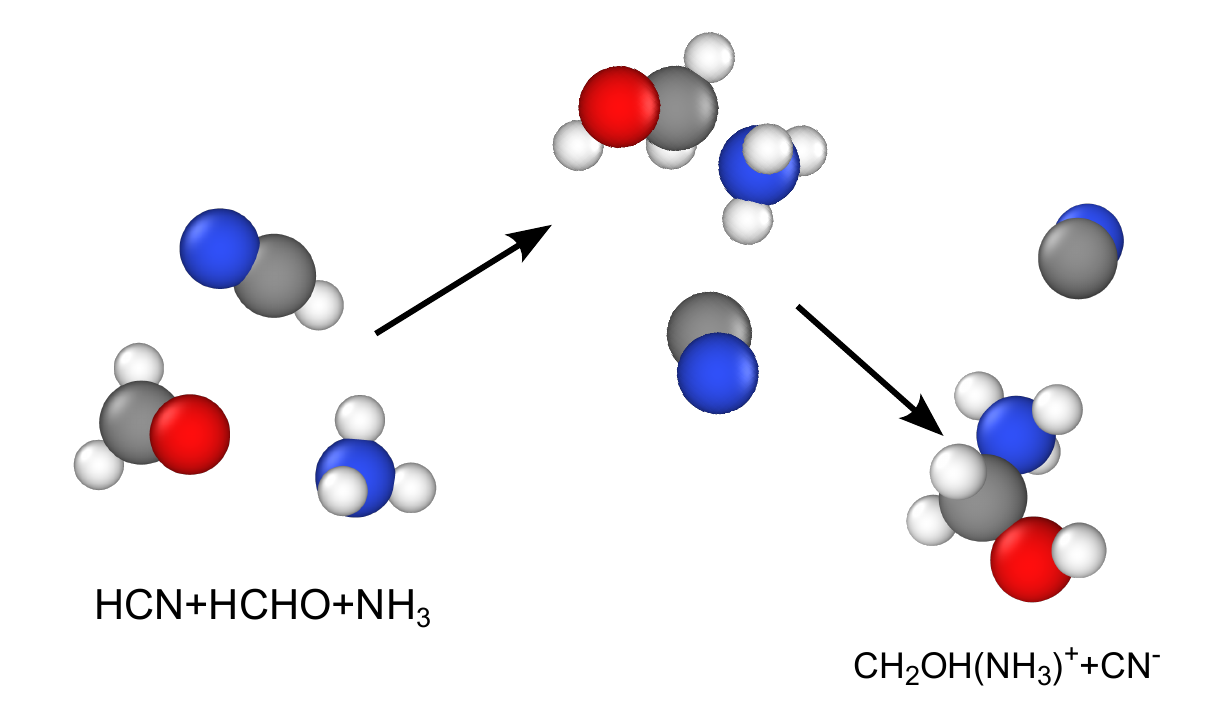}
    \caption{First step of the Strecker synthesis of glycine investigated in this work (cf. Fig.~\ref{fig:full_strecker}). The reaction starts with a proton transfer from hydrogen cyanide to formaldehyde resulting in the formation of the transition state. In the next step, the addition of ammonia leads to the formation of protonated aminomethanol as intermediate product. Hydrogen, carbon, nitrogen and oxygen atoms are colored in white, grey, blue and red, respectively. The surrounding water molecules are not shown for clarity.}
    \label{fig:part_reaction}
\end{figure}

\begin{figure}
    \centering
    \includegraphics[width=\columnwidth]{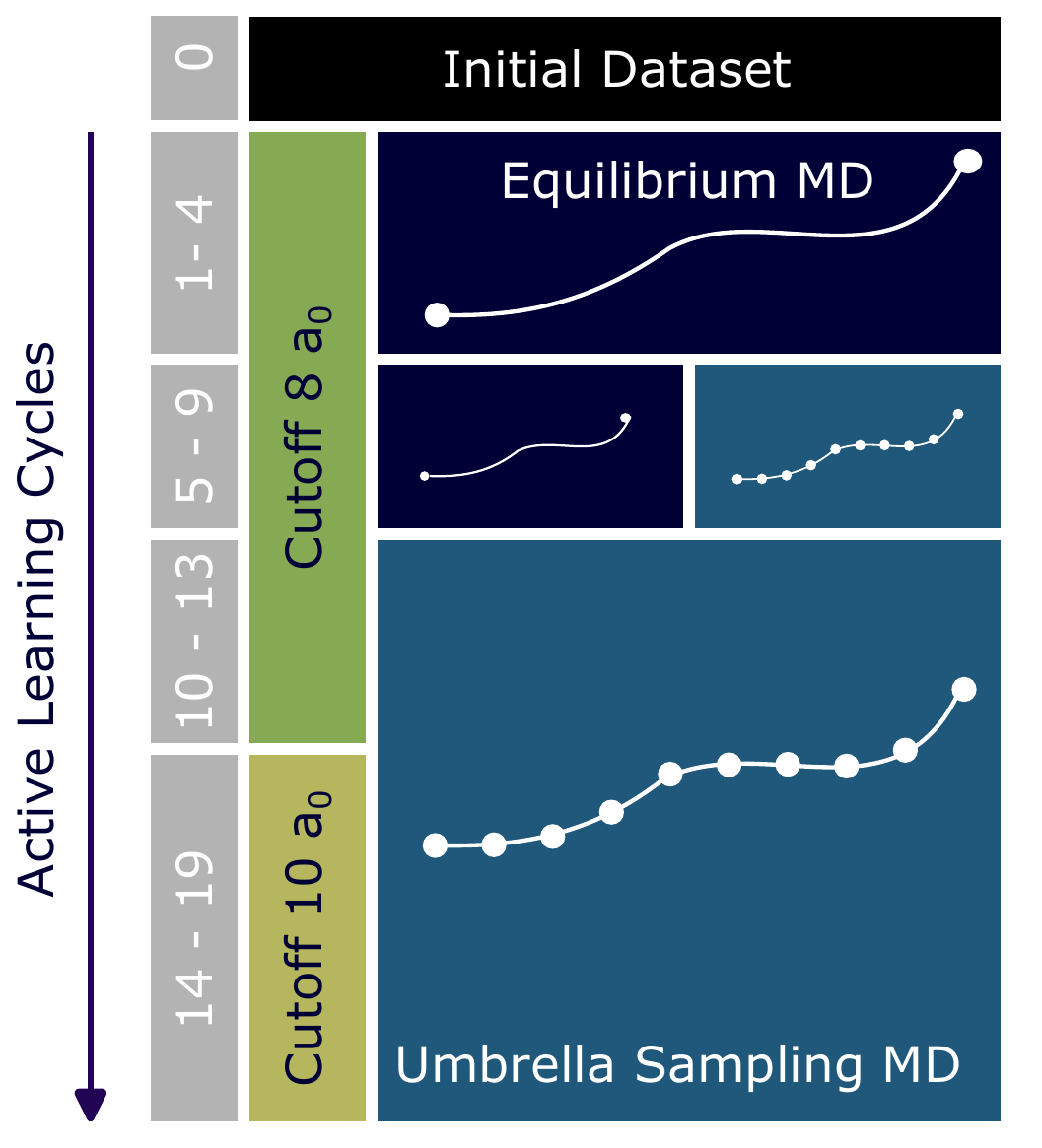}
    \caption{Schematic workflow of the iterative extension of the reference dataset by AL employed in the present work. The process starts with a small initial dataset and a cutoff radius of $8\,\mathrm{a_0}$ for constructing preliminary HDNNPs. In the first four cycles, this dataset is extended by sampling at the initial and final equilibrium states by unbiased MD. Starting in cycle five, in addition umbrella sampling simulations along the reaction path are performed. After 9 cycles only the umbrella sampling simulations are continued to identify new structures. Finally, in cycle 14 the cutoff radius is increased to $10\,\mathrm{a_0}$ to refine the description of the atomic environments in the final potentials.
    }
    \label{fig:dataset_construction}
\end{figure}

The starting point for constructing the reference dataset is the reaction path of the first step of the Strecker synthesis (cf. Fig.~\ref{fig:part_reaction}), which has been mapped by metadynamics simulations and committor analysis in previous work~\cite{magrino_step_2021}. Of this data 55 structures have been selected which are the initial configurations of the umbrella sampling simulations.  
Moreover, 1212 configurations have been taken from the available AIMD trajectories of the initial and the final equilibrium states at 300~K, yielding in total 1267 structures in AL cycle 0. These structures have been recomputed by DFT employing the setup described below to obtain reference energies and forces.

Since our goal is the construction of a reference dataset without running demanding AIMD trajectories in each umbrella sampling window, this first reference dataset has then been extended by AL as described in Ref.~\citenum{P6057,eckhoffMolecularFragmentsBulk2019}.
In general, AL is an iterative procedure in which the dataset is extended by new configurations chosen from a pool of candidate structures generated, e.g., in MD simulations using preliminary MLPs\;\cite{P3114,liebetrau_hydrogen_2024, eckhoffInsightsLithiumManganese2021a, eckhoffClosingGapTheory2020, daru_coupled_2022,kapil_first-principles_2022,schran_machine_2021, chen_data-efficient_2023, deringer_machine_2017,schran_automated_2020, schran_committee_2020, tang_machine_2023,P5782}. Due to the flexible functional form of MLPs their predictive accuracy is limited for configurations that are very different from the reference data set resulting in a prediction uncertainty for structures not well represented by the current dataset.
This uncertainty, or variance, can then be used to decide which structures should be added to the reference dataset. Also other criteria and strategies to sample new structures have been proposed~\cite{raff_ab_2005,doughan_theoretical_2006,zhang_modeling_2023,stolte_random_2024, benayadPrebioticChemicalReactivity2024, P4939,guo_checmate_2023}.

In this work we extend the initial dataset in 19 cycles of AL. In each cycle we train six HDNNPs to the current dataset and use the best two potentials to propagate MD simulations, details of the fitting process are given in Sec.\;\ref{subsec: HDNNP_construction}. Simulations are terminated at a maximum length of 100\,ps or are stopped when a threshold of accumulated extrapolation warnings is reached (see Sec.\,\ref{subsec:MDval}). 

As we need to sample reaction barrier crossings to construct a reference data set, we enhance sampling configurations at this energetically less favorable region by employing umbrella sampling. This enables to collect candidate structures in several HDNNP-based MD trajectories in confined windows along the reaction path. Of this pool of candidate structures, new structures are identified based on the variation of the energy or force component prediction of the two best HDNNPs. The required threshold has been set between $0.08-0.005\;\mathrm{eV/atom}$ for energies and between $0.65-0.35$~eV/a$_0$ for force components. The thresholds are decreased with advances in the AL such that per trial simulation usually one and at most two new configurations are chosen.
After identifying in total a few hundred new structures, they are recalculated with DFT and added to the dataset. Then, the potentials are retrained and a new cycle of AL is started.

Since the initial dataset covers only a part of the target configuration space, the AL process has been started employing a relatively small ACSF cutoff radius of 8 Bohr. This provides a more robust fit which increases the stability of early simulations compared to an HDNNP with same number of symmetry functions and a larger cutoff radius. Moreover, using a smaller cutoff speeds-up the simulations and the training. However, when employing a small cutoff, relevant  interactions may be missing ultimately limiting the accuracy that can be obtained. Thus, the smaller cutoff radius is only used in the initial phase of AL. In the final cycles it has been increased to 10 Bohr and additional ACSFs were introduced which are described in more detail in Sec.\;\ref{subsec:HDNNP_constr}. 

The different phases of AL are summarized in Fig.\;\ref{fig:dataset_construction}. The procedure is started with using unbiased MD simulations at the initial and final equilibrium states with sampling temperatures 200, 250, 300, and 350~K. Once pure solvent simulation reached reasonable reliability, as discussed in Sec.\;\ref{sec:results_discussion}, umbrella sampling AL was additionally employed at temperatures of 200, 250, 300, 350, and 400~K. After in total 9 AL cycles only umbrella sampling simulations were employed to generate new structures. Information about the number of simulations, which were performed in each cycle, is given in the SI.

\subsection{Density functional theory calculations}

The reference DFT calculations to determine the energies and forces of the training and test structures were carried out using the Fritz-Haber-Institute ab initio molecular simulations (FHI-aims) program package \cite{blumInitioMolecularSimulations2009a} (version 221103) employing a numerical atomic orbital basis.
The RPBE functional\;\cite{hammer_improved_1999} has been chosen to describe exchange and correlation in combination with D3 dispersion corrections~\cite{grimmeConsistentAccurateInitio2010a} (DFT-D3 program version 3.1 Rev 1 of October 2015) using zero damping. This has been shown to provide a reasonable description of the properties of liquid water~\cite{P4556}. “Intermediate” settings have been employed for the integration grid and the basis sets used to expand the Kohn-Sham orbitals. A 2$\times$2$\times$2 k-point grid was chosen for all calculations.
The convergence criterion for electronic self-consistency of the single point calculations has been set to $10^{-6}\,\mathrm{eV}$ for the total energy and $10^{-4}\,\mathrm{eV/\AA}$ for the forces. 

The system (cf.~Fig.~\ref{fig:part_reaction}) contains the reactant species and 80 water molecules resulting in total in 251 atoms that are placed in a periodic cubic box of $13.4\,\mathrm{\AA}$-side length. 
For studies of pure water, an orthorhombic box of size $16.5\,\mathrm{\AA} \times 16.5\,\mathrm{\AA} \times 17.0\,\mathrm{\AA}$ containing 160 water molecules has been used. 

\subsection{Construction of the high-dimensional neural network potentials} \label{subsec: HDNNP_construction}
\label{subsec:HDNNP_constr}

The HDNNPs have been parameterized using the RuNNer code~\cite{P4444,behlerFirstPrinciplesNeural2017}. For each AL cycle six different HDNNPs have been trained employing two different random seeds for the initial weight parameters as well as three different atomic NN architectures. These architectures consist of two hidden layers containing 30 and 25 nodes, three hidden layers containing 25, 20, and 15 nodes, and three hidden layers containing 20, 15, and 10 nodes, respectively. For a given HDNNP, the same architecture has been used for all elements. The weights of the atomic NNs with two hidden layers were initialized with the scheme proposed by Nguyen and Widrow\;\cite{nguyenImprovingLearningSpeed1990} and preconditioned to give an energy distribution of same mean and standard deviation as the reference energy distribution~\cite{P4444}. The weights of the the atomic NNs with three hidden layers were initialized following the method of Xavier\;\cite{xavierglorotUnderstandingDifficultyTraining2010} including a modification proposed by Eckhoff et al.\;\cite{eckhoffClosingGapTheory2020}.

The parameters of the ACSFs have been automatically determined and adapted with respect to the increasing structural diversity in the dataset during AL. Initially, a small cutoff of $8\,\mathrm{a_0}$ has been employed, where $a_{\mathrm{0}}$ is the Bohr radius. This cutoff has been extended to $10\,\mathrm{a_0}$ in AL cycle 14 to refine the structural description in the final stage of AL. For the larger cutoff, a second set of angular symmetry functions has been included. The details of the ACSFs and their parameter values are given in the SI. For training the HDNNP, the values of each ACSF have been scaled to the range of [-0.5,0.5].
The weights of the atomic NNs were iteratively optimized for 20 epochs to minimize the total energy and atomic force errors employing the global extended Kalman filter~\cite{blankAdaptiveGlobalExtended1994}. 90\% of the reference structures have been used to train the HDNNP and 10\% were kept for testing the generalization ability to structures not included in the training set. In each iteration, 1 \% of the force components have been randomly chosen for updating the weights to speed-up the training process, while all total energies have been used. 
Each force update was followed by a repeated energy update to ensure a balanced number of energy and force updates. 
The adaptive threshold of the global, extended Kalman filter\,\cite{P0425,P1308,P4610} was set to a factor of $0.5$ of the current RMSE for energies and forces and was increased to a value of $0.8$ in AL cycle 15. 
After training, in each AL cycle the two HDNNPs with lowest test set energy and force root mean squared errors (RMSEs) have been selected to extend the reference dataset.

\subsection{Molecular dynamics simulations}
\label{subsec:MD}
The MD simulations were performed utilizing the Large-scale Atomic/Molecular Massively Parallel Simulator (LAMMPS, version 2 Aug 2023)\,\cite{LAMMPS}, including the n2p2 library for HDNNPs (version 2.2.0)~\cite{LibraryBasedLAMMPSImplementation}. 
All MD simulations were run in the canonical $NVT$ ensemble at 300\;K. A time step of $\delta t= 0.5\,\mathrm{fs}$ was employed with a hydrogen mass of $2$\;u as in previous work~\cite{magrino_step_2021}. MD simulations of pure water were performed with a timestep of $\delta t=0.25\;\mathrm{fs}$ using a hydrogen mass of $1.008$\;u. The Nos\'e-Hoover thermostat\,\cite{evansNoseHooverThermostat1985} was applied with a damping parameter of 100 times the timestep. The velocity Verlet algorithm\,\cite{swopeComputerSimulationMethod1982a} was chosen as integrator for the equations of motion. During all MD simulations the ACSF values of all atomic environments have been monitored and in case a value outside the range covered by the reference data set has been encountered, an extrapolation warning has been issued for further analysis.

The umbrella sampling simulations were conducted with LAMMPS using the open-source, community-developed PLUMED library (version 2.8.2)~\cite{bonomiPromotingTransparencyReproducibility2019, tribelloPLUMEDNewFeathers2014}.
The overall simulation setup was taken from Ref.~\cite{magrino_step_2021}.
Specifically, the element-pair dependent parameters $r_0^{\alpha \sigma}$ for the calculation of the atomic coordination numbers were set to $1.4$~\AA{} for hydrogen/carbon and nitrogen/oxygen pairs and to $1.8$~\AA{} for all other element pairs. In total $P=12$ configurations including two structures for the reactant and product equilibrium state were chosen to define the reaction path for the calculation of the CVs $s$ and $z$. In the umbrella sampling simulations the system was confined to 55 windows along the $s$ path CV with a umbrella sampling bias potential of strength $k = 1.74153\;\mathrm{eV}$. In addition, the $z$ path CV was restricted by a semiparabolic wall at $z=0.12$ with $k=100\;\mathrm{eV}$ and a semiparabolic wall with $k=50\;\mathrm{eV}$ was applied directly to the coordination of the aldehyde carbon atom by nitrogen atoms and the coordination of the cyanide carbon atom by hydrogen atoms to improve the confinement of the system to the reaction path and to avoid any unwanted hysteresis effects.

For the reweighing of the biased simulations and the calculation of the free energy as given in Eq.\;\ref{eq:free_energy} WHAM~\cite{kumarMultidimensionalFreeEnergy1995} was used as implemented in the Grossfield code~\cite{GrossfieldLabConnecting}. The convergence criterion of WHAM was set to $10^{-7}$ kcal/mol and 150 bins in $s$-space were employed. 
The statistical uncertainty of the free energy profile has been estimated by comparing the free energy profiles obtained from the third and fourth quarters of each umbrella sampling trajectory. 

\section{Results and Discussion} \label{sec:results_discussion}

\subsection{Reference dataset} 
\label{subsec:datasets}

\begin{figure}
    \centering
    \includegraphics[width=\columnwidth]{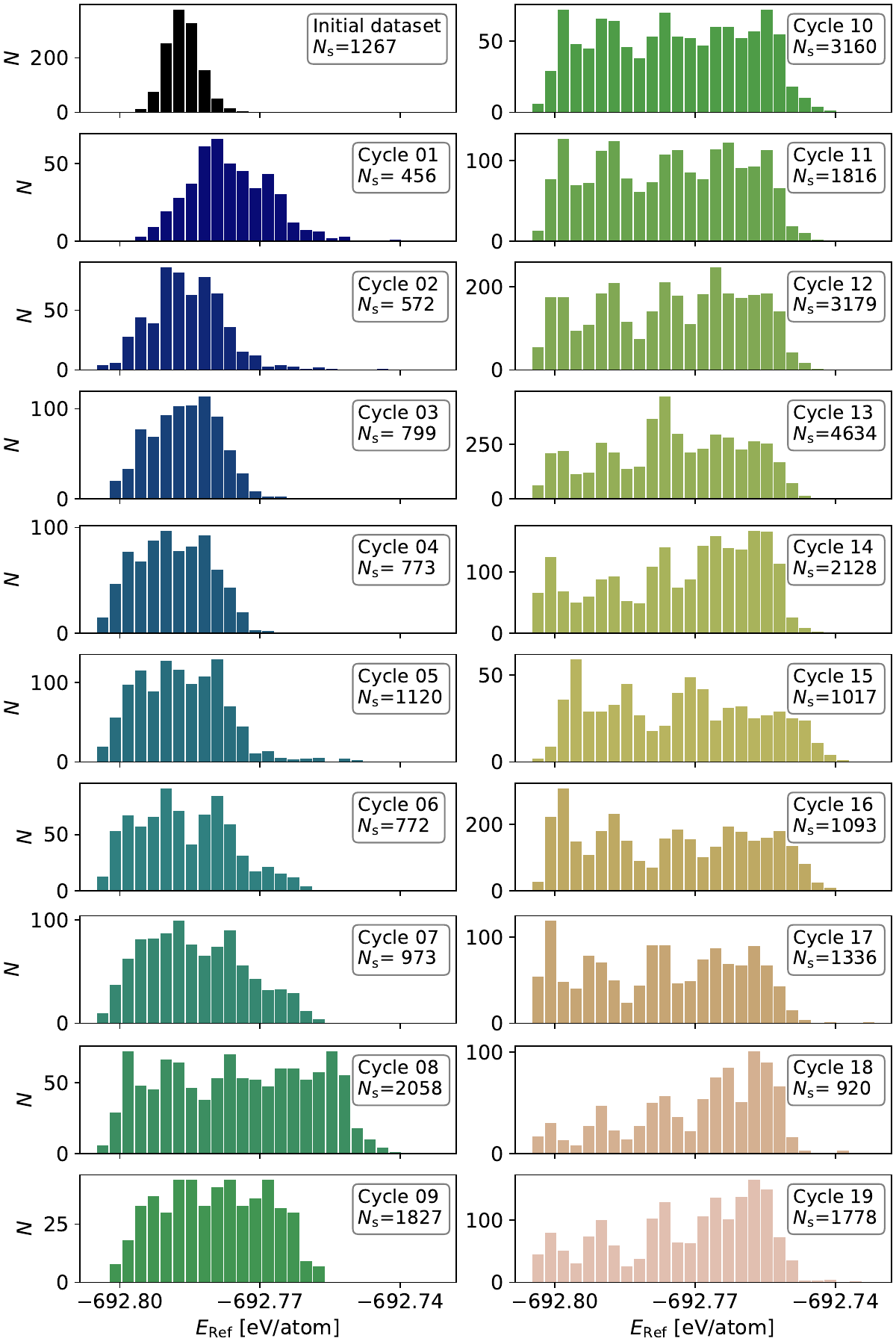}
    \caption{Distributions of the DFT energies $E_{\mathrm{Ref}}$ for the $N_\mathrm{s}$ reference structures added in each of the 19 AL cycles. $N$ is the number of structures per histogram bin. Compared to the initial dataset obtained from AIMD at 300~K, the energy distributions of the structures added by AL are broader due to the increased range of sampling temperatures up to 400~K. }
    \label{fig:energy_distribution}
\end{figure}

The DFT reference dataset has been constructed iteratively by AL employing MD and umbrella sampling simulations as described in Section \ref{subsec:dataset_construction}. In total, a large number of 19 AL cycles has been carried out to enable detailed convergence tests with respect to the dataset size. 
The final dataset consists of 31,678 structures which cover all umbrella sampling windows. It contains approximately 60,000 atomic environments of carbon and nitrogen atoms, about 2,500,000 oxygen atomic environments and roughly 5,000,000 hydrogen atomic environments.

The energy distributions in the subsets of structures added during the AL cycles are shown in Fig.\;\ref{fig:energy_distribution}. The energy range of the initial dataset obtained from equilibrium AIMD simulations at 300\;K is rather narrow compared to the energy distributions of the datasets obtained by the subsequent AL cycles. The reason for this broadened energy range is twofold. First, a wider range of temperatures from 200~K up to 400~K has been used in AL to sample structures of increased diversity. Further, 
the selection of structures by AL biases the added data points to atomic configurations, which are underrepresented in equilibrium MD at room temperature and are likely to exhibit slightly above average potential energies. 

Investigating the energy distributions in Fig.~\ref{fig:energy_distribution} more closely, it can be observed that in the initial AL phase consisting of cycles 1-4 employing only unbiased MD the mean energy of the distributions first shifts to higher values (cycles 1 and 2). Then, the centers of the distributions decrease again to align with the center of the initial dataset (cycles 3 and 4). Apart from the broader temperature range, the main reason for the initial increase is the limited coverage of energetically higher parts of configuration space by the AIMD data of cycle 0. As a consequence, the energy and force predictions of less well represented structures, e.g. repulsive structures containing shorter interatomic distances, are unreliable and can guide the simulations to energetically less favorable geometries. These exhibit a high committee uncertainty and are thus selected by the AL algorithm. This finding is in good agreement with previous studies of Stolte et al. for water\;\cite{stolte_random_2024}. Once the HDNNP learns these atomic interactions, the simulations become more stable, avoid too high-energy regions and sample structures in an energy range more similar to those visited in AIMD simulations. 

From cycle 5 onward, umbrella sampling simulations along the reaction path are used in the AL process, and structures covering a broad range of energies are added to the dataset as new parts of configuration space along the reaction path are explored. As discussed below (cf. Section~\ref{subsec:MDval}), this leads to a continuous improvement of the stability of the simulations resulting in longer trajectories further improving the sampling. Thus, the fraction of higher energy structures added to the dataset remains high and becomes even dominant in the final AL cycles, in which only a few low-energy structures in the well-covered region are still found.\\ 

\begin{figure}
    \centering
    \includegraphics[width=\columnwidth]{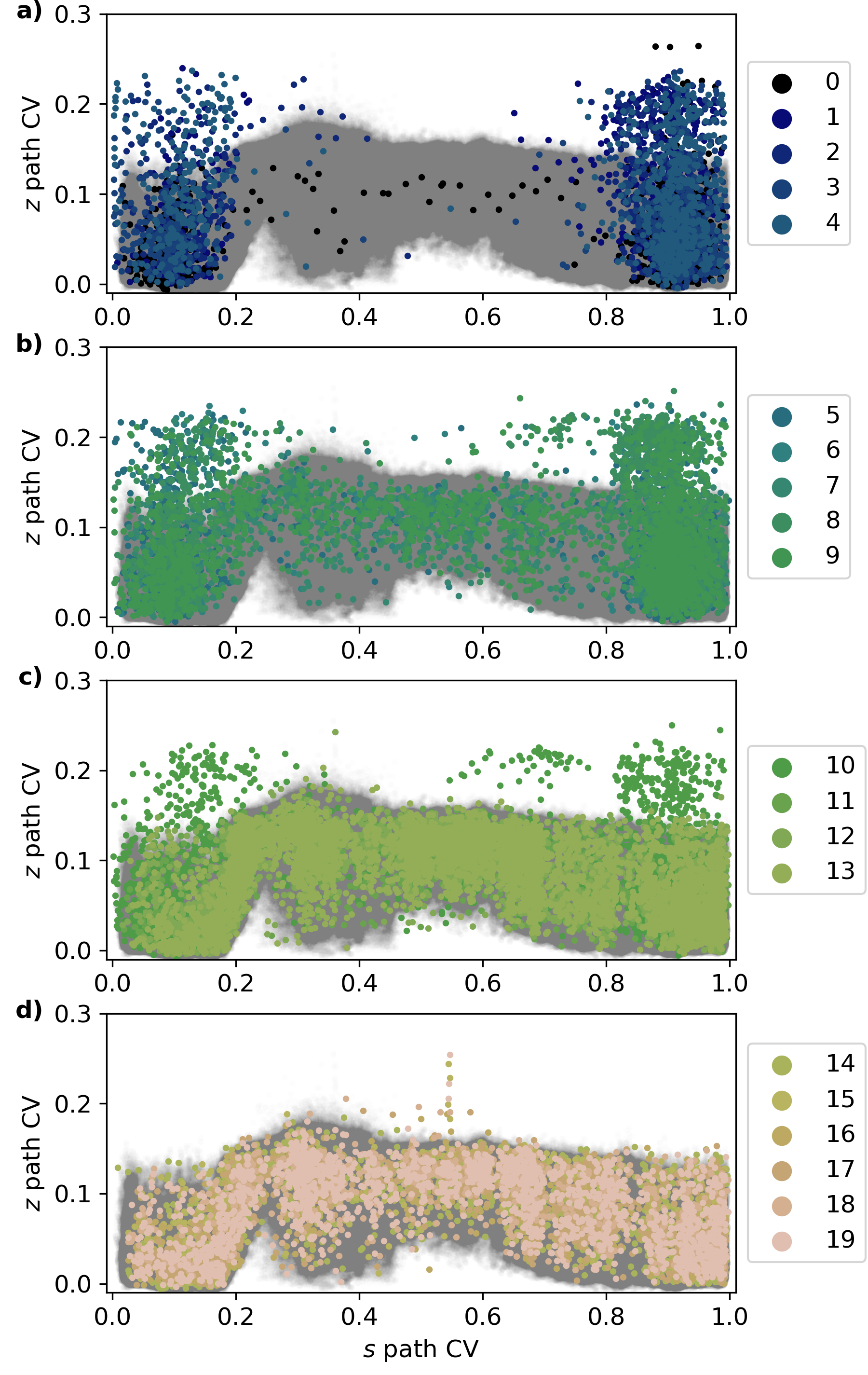}
    \caption{ Distribution of the reference data points in the ($s,z$) path CV space. The colors correspond to the AL cycle in which the points have been added.  Panel (a) shows the initial dataset and the points resulting from AL using MD only at the initial and final equilibrium states in cycles 1-4, as well as one point included for each window along the reaction path. In cycles 5-9 umbrella sampling and MD simulations are used to identify generate new structures along the reaction path (b). Panel (c) shows the data of cycles 10-13, which only use umbrella sampling, before finally the cutoff is increased to $10\,\mathrm{a_0}$ for the final cycles in panel (d). The grey area represents the full configuration space visited during umbrella sampling simulations of the reaction using the final converged potential obtained in cycle 19.}
    \label{fig:dataset_sz}
\end{figure}

Figure\;\ref{fig:dataset_sz} shows the exploration of configurations in the ($s$,$z$) path CV space  along the reaction path during AL.  The data has been grouped in four panels according to the four phases of the AL procedure (cf. Fig.\;\ref{fig:dataset_construction}). For comparison, also the distribution of the $s$ and $z$ values of the structures visited in extended umbrella sampling simulations performed using a converged HDNNP obtained in AL cycle 19 at 300\;K is provided as a grey-shaded area to highlight the region in ($s$,$z$) space that is required for the final applications.

The CV values $s=0.1$ and $s=0.9$ correspond to the reactant and product equilibrium states of the reaction, and the AL procedure starts with equilibrium MD simulations for these states as well as a single structure for each umbrella sampling window (Fig.\;\ref{fig:dataset_sz}a). The next phase of AL covering cycles 5 to 9 additionally uses umbrella sampling simulations to sample new configurations along the reaction path (Fig.\;\ref{fig:dataset_sz}b). 
It can be observed that at the beginning of the AL process preliminary HDNNPs tend to generate structures outside the ($s,z$) path CV space that is visited in the reference umbrella sampling simulations using the converged, i.e., fully reliable, potential. As the AL process continues (Figs.~\ref{fig:dataset_sz}c and d) the amount of these outliers decreases significantly and new selected data points are mainly sampled in the relevant part of the ($s$,$z$) space. However, it should be noted that due to the rather crude characterization of the structures by the ($s,z$) path CVs there is no direct correlation between the spatial proximity of a point to the reaction path and its potential energy as shown in Fig.\;S1 in the SI 

Further, the density of the points selected by AL is not equally distributed along the reaction path. For instance, around the transition state at approximately $s=0.4$ a small gap is present, which turns out to be difficult to sample. On the other hand, the densities of points in the reactant and in particular in the product basins remain high, which indicates that also here new geometries are found, which, however, often have relatively high potential energies that are less important for MD simulations at 300~K (cf. Fig.\;\ref{fig:energy_distribution}).\\

\begin{figure*}
    \centering
    \includegraphics[width=2\columnwidth]{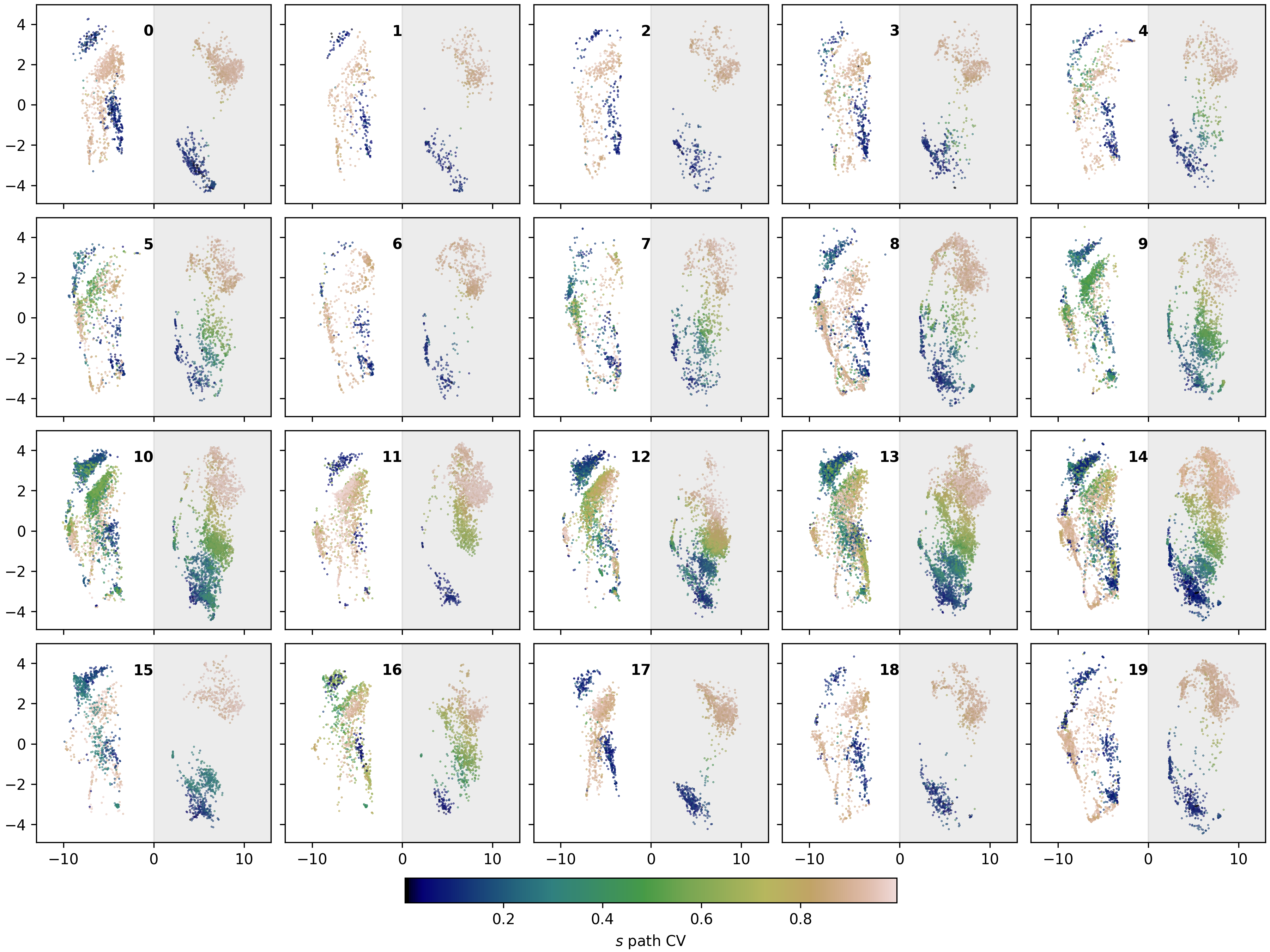}
    \caption{Two-dimensional visualization of the carbon atomic environments in the ACSF space for the structures added in each AL cycle obtained from t-SNE dimensionality reduction. The x and y-axis are the first and second dimension of the ACSF vector in the two-dimensional reduced space. For comparison, all environments are characterized using the same ACSF set of cycle 19. Points are colored by the $s$ value of the structure of the respective atomic environment. t-SNE clearly separates atomic environments of the carbon atom in formaldehyde and the carbon atom in cyanide in two distinct clusters which are located in different regions with grey and white background, respectively. The number of the AL cycle is given in each panel. 
}
    \label{fig:dataset_water_dim}
\end{figure*}

While the distribution of data points in the path CV space provides valuable information about the mapping during AL along the reaction pathway, it is also of high interest to analyze the increasing structural diversity of the atomic environments in the course of AL. This diversity is difficult to analyze in the high-dimensional ACSF space characterizing the atomic environments directly, but it can be visualized employing dimensionality reduction techniques such as t-distributed stochastic neighbor embedding (t-SNE)\;\cite{NIPS2002_6150ccc6, maatenVisualizingDataUsing2008} embedding the high-dimensional ACSF vector of each environment in two dimensions. As an example, Fig.\;\ref{fig:dataset_water_dim} shows two-dimensional representations of the local environments of both carbon atoms in the system in all AL cycles. Since the parameters of the ACSFs in the simulations change with increasing dataset, for comparison in this analysis we use the same final set of ACSFs of cycle 19 for all data points of all cycles. As can be seen, t-SNE assigns the atomic environments in two distinct clusters separating the environments of the carbon atom of formaldehyde and the carbon atom of cyanide. The points of the formaldehyde carbon environments cluster form child clusters, which are especially well separated in cycle 0. These child clusters can be assigned to reactant and product environments. The clear separation in child clusters can be expected due to strong structural changes induced by the reaction. In the progress of the AL new environments between the reactant and product cluster are found.
The cyanide carbon environments change less during the reaction resulting in less pronounced subclustering. 
In different AL cycles different geometric environments are selected and umbrella sampling simulations along the reaction path are needed to cover all relevant configurations.

A similar visualization of the environments of the oxygen atoms can be found in Fig.~S2 in the SI. In this case, the changes in the t-SNE plot are less pronounced after AL cycle 4 indicating that the solvent sampling is essentially completed in the initial phase of AL.

\subsection{Accuracy of the HDNNP}
\begin{figure}
    \centering
    \includegraphics[width=\columnwidth]{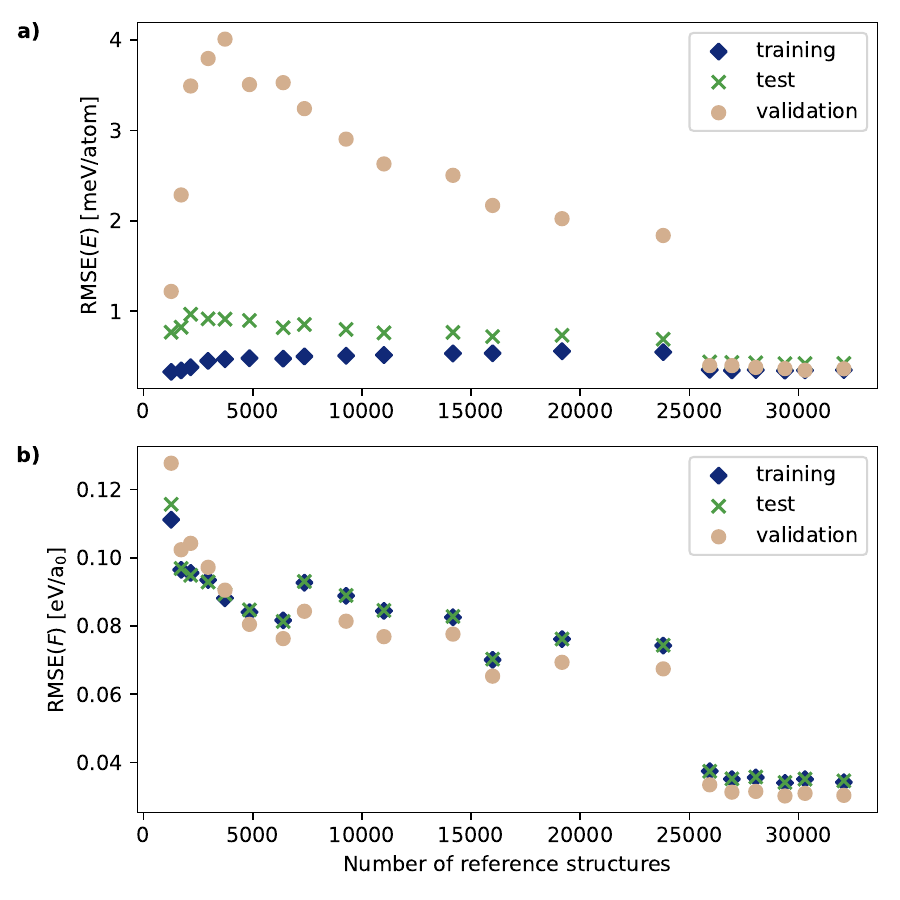}
    \caption{Learning curves showing the RMSEs of training sets (90~\% of reference structures) and test sets (10~\% of reference structures) of increasing size as well as for a fixed validation set for the energies (a) and the force components (b) during the AL process. 
    The validation set, which covers the full configuration space, contains the DFT data of 220 structures extracted from umbrella sampling simulations of all windows obtained with a HDNNP of AL cycle 19. The significant decrease in the RMSEs beyond 25,000 structures, i.e., in the last six AL cycles, is due to the increase of the cutoff radius from 8 to 10 $\mathrm{a}_0$ and the introduction of a second set of angular ACSFs. }
    \label{fig:learning_curve}
\end{figure}

\begin{figure}
    \centering
    \includegraphics[width=\columnwidth]{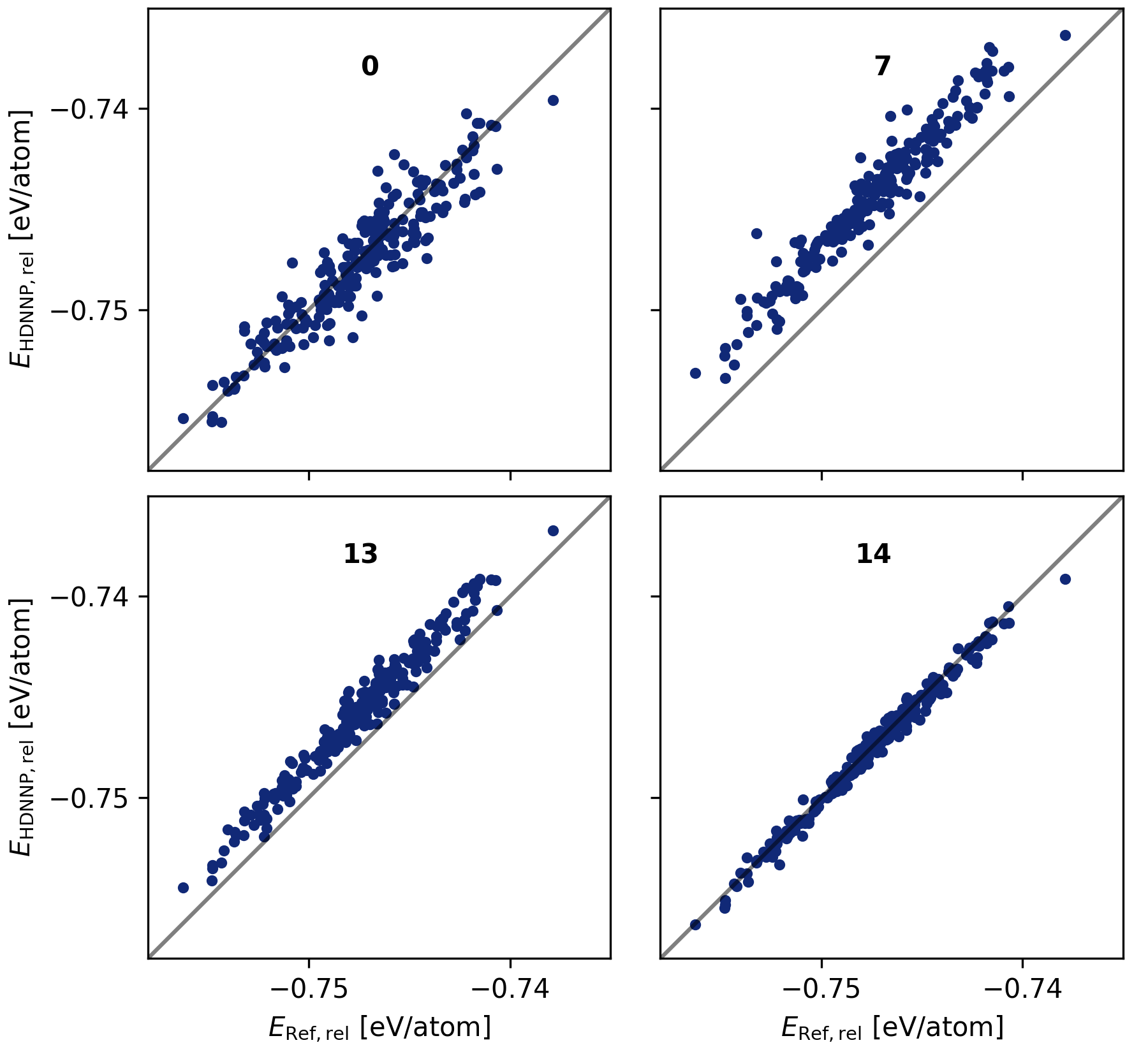}
    \caption{Correlation of HDNNP energies and reference DFT energies of the validation set in selected AL cycles. Shown are relative energies with the same offset for both axes. The respective AL cycle is given in each panel.
    }
    \label{fig:val_corr}
\end{figure}

In order to monitor the improvement in the accuracy of the energies and forces during the extension of the dataset by AL the RMSEs of training and test sets of increasing size as well as of a fixed validation set are shown in the learning curves in Fig.\;\ref{fig:learning_curve}. For each dataset size, i.e., AL cycle, the test set consists of 10~\% of the structures from the pool of reference structures, which have been randomly selected, and the training set contains the remaining 90~\%. The validation set consists of the DFT data of 220 structures extracted from all 55 windows of umbrella sampling simulation obtained in AL cycle 19 covering the full relevant configuration space. For consistency, the same NN architecture (two hidden layers with 30 and 25 neurons respectively) providing overall the smallest force and energy RMSE of the test set has been used for all AL cycles.

Overall, the training and test set RMSEs in all cycles are very low, i.e., below 1~meV/atom for the energies, and -- with the exception of the smallest dataset -- lower than 0.1~eV/a$_0$ for the force components. Further, the typical shape of learning curves can be observed in Fig.~\ref{fig:learning_curve}a. While the error of the training energies slightly increases with growing complexity of the dataset, at the same time the generalization capabilities of the HDNNP improve resulting in smaller energy errors of the test set structures. 
This relation holds true for most of the test set errors. However, at the beginning an initial rise in the test set RMSE of the energies is observed, which is a consequence of the addition of higher energy structure in the early AL cycles, which increase the complexity of the configuration space to be learned.

The training and test set errors of the force components are very similar already in an early stage of the learning curves, which has two reasons. First, the number of force components available in the dataset is much larger than the number of energies, which improves the fit quality already in the early AL cycles. Second, to reduce the computational costs of the force training only a random subset of the available training forces is used per epoch. Consequently, only a part of the available information is used for training, which slightly increases the training set force RMSE compared to when using all force components of the training set. Still, as this subset is different in each epoch, the alternating use of different force components during training ensures a good coverage of the overall PES such that the test forces are not very different from the training forces resulting in similar RMSEs.

During AL, the force RMSEs decrease further with increasing dataset size.
Both, energy and force errors reach a plateau when about 20,000 structures are included, indicating convergence of the accuracy of the potential for the initial ACSF cutoff of 8~a$_0$. Then, in the final six AL cycles using the increased cutoff of 10~a$_0$ and a second set of angular ACSFs, both the energy and force errors with respect to DFT of training and test set  are further substantially reduced to approximately half of the values obtained with a cutoff of 8~a$_0$.
However, during these cycles also the errors obtained with the larger cutoff do not significantly decrease further with the addition of more structures to the dataset, which demonstrates that also the larger configuration space of the extended atomic environments is well covered.

The final HDNNP in AL cycle 19 has an energy RMSE of $0.35\;\mathrm{meV/atom}$ for the training set and $0.42\;\mathrm{meV/atom}$ for the test set. The RMSE of the force components is $34.2\;\mathrm{meV/a_0}$  for training set and $34.6\;\mathrm{meV/a_0}$ for the test set. These errors are very low and in the typical order of magnitude of current state-of-the-art MLPs.

It is important to note that the training and test sets in Fig.~\ref{fig:learning_curve} have been constructed from the reference data sets available at the respective AL cycles. Therefore, in particular in early AL cycles they do not cover the full configuration space along the reaction path. To obtain a realistic assessment of the quality in the description of all relevant configurations in all AL cycles, we have constructed in addition a validation dataset covering the full reaction path using DFT energies and forces computed for four structures from each of the 55 umbrella sampling windows which have been obtained with the final HDNNP at 300~K. 
The energy and force RMSEs of this validation set containing 220 structures is also included in Fig.~\ref{fig:learning_curve}. As expected, the energy error is generally higher than the error of the test set, since the test set is confined to the same reduced configuration space as the training set, while the validation set is the same for all AL cycles.

Several interesting observations can be made. First, the energy error in the early AL cycles is surprisingly low, which is likely a consequence of the same sampling temperature, i.e., potential energy range, in the early training sets and the validation set. More surprisingly, in cycle 13 the validation energy error saturates at a relatively high error of about 2~meV/atom, which after increasing the cutoff radius to 10~a$_0$ in cycle 14 immediately decreases to the very low errors of the training and test sets. To further analyze this observation, Fig.~\ref{fig:val_corr} shows the energy correlation plots of the validation set with respect to DFT for selected AL cycles. It can be clearly seen that the reason for the large RMSEs of the validation set up to cycle 13 is a systematic energy offset. This offset is caused by long-ranged interactions, which are included only in an average way if a small cutoff of 8~a$_0$ is used, and which seems to be emerging due to the inclusion of an increased number of higher energy structures. Similar observations have been made by Stolte et al. \cite{stolte_random_2024} for the case of liquid water for the case of biases in the selection of structures through AL. As soon as the cutoff is increased to 10~a$_0$ in cycle 14, these interactions can be captured by the improved description of the atomic environments resulting in a drastically improved accuracy of the potential energies. 
Since total energy offsets are not relevant for energy gradients, the RMSEs of the forces of the validation set shown in Fig.~\ref{fig:learning_curve}b are very similar to the respective values of the training and the test set for all AL cycles. Still, the force RMSEs of the validation set are generally slightly lower than for the training and test sets, since the validation set is restricted to configurations relevant for simulations at 300~K, while the training and test set also include higher energy structures. 
Moreover, a general improvement of the forces with an increase of the cutoff to 10~a$_0$ is found. These data clearly support our finding that a large cutoff is needed for an accurate description of this system.\\

\begin{figure}
    \centering
    \includegraphics[width=\columnwidth]{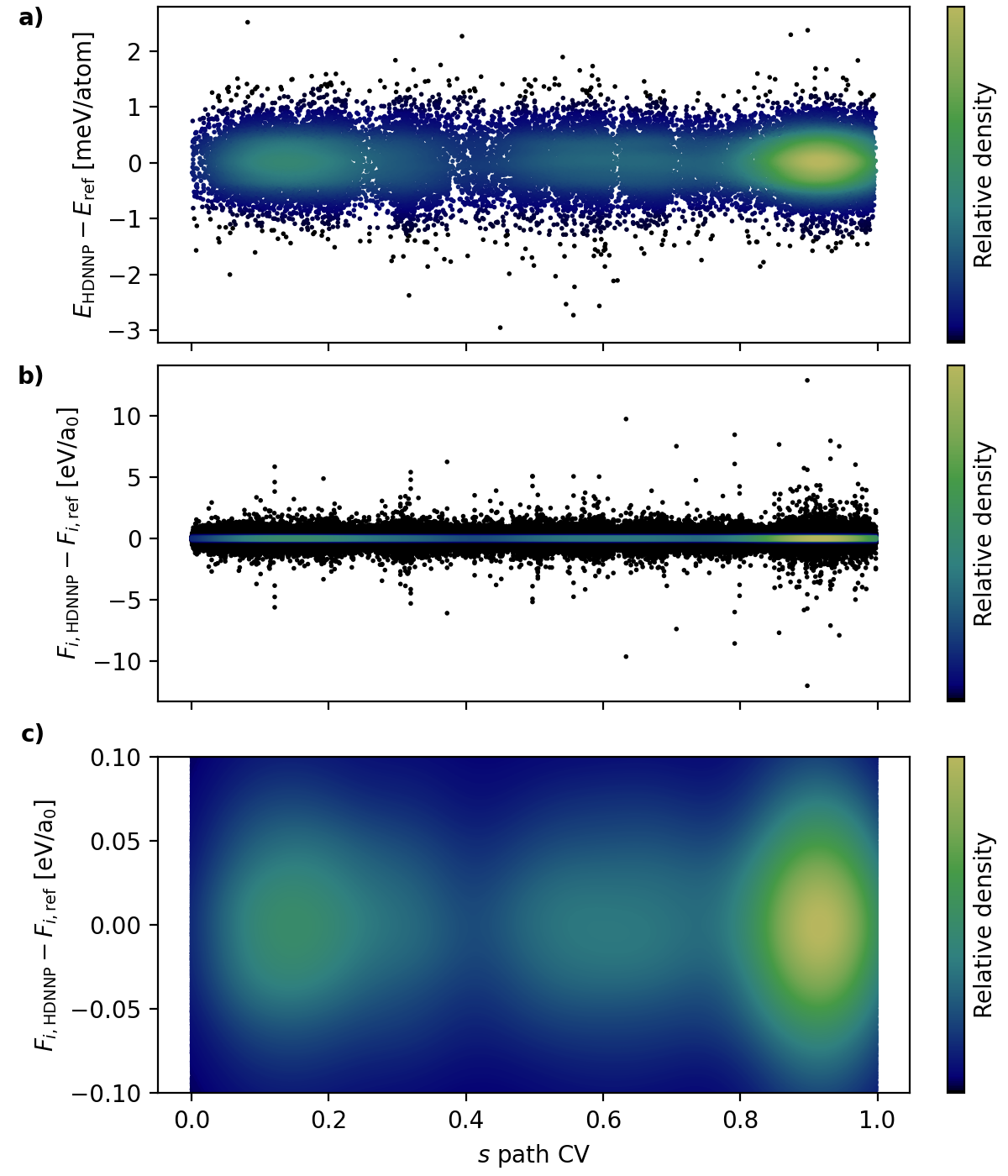}
    \caption{Distributions of the energy (a) and force component (b) errors between the HDNNP and the DFT reference calculations (training and test set) along the reaction path CV $s$ in AL cycle 19. Panel (c) shows a zoom of the force component errors in the central region of panel (b). The colors represent the relative density of points.  
    }
    \label{fig:DE_vs_s}
\end{figure}

While obtaining low RMSEs of energies and forces is a mandatory condition for a reliable potential, they are not sufficient to assess the reliability of the potential as they do not provide direct information about the distribution of errors and possible outliers in the dataset.
Figure\;\ref{fig:DE_vs_s} gives an overview of the energy and force errors of all reference structures along the reaction pathway. As can be seen, most of the energy errors are in the interval between -0.5~meV/atom and 0.5~meV/atom and in the range between -0.05~eV/a and 0.05~eV/a$_0$ for the force component error for all $s$ values suggesting a high accuracy of the HDNNPs for all structures along the reaction path. Only a few marginal outliers with energy deviations up to about $\pm$2 meV/atom are present (Fig.~\ref{fig:DE_vs_s}a), and for the force components only about 4,600 out of in total 24,000,000 components exhibit errors larger than 1~eV/a$_0$ (Fig.~\ref{fig:DE_vs_s}b). 
A visualization of errors of all energies and forces are given separately for the training and the test set for all AL cycles in Figs.~S4-S43.\\

\begin{figure}
    \centering
    \includegraphics[width=\columnwidth]{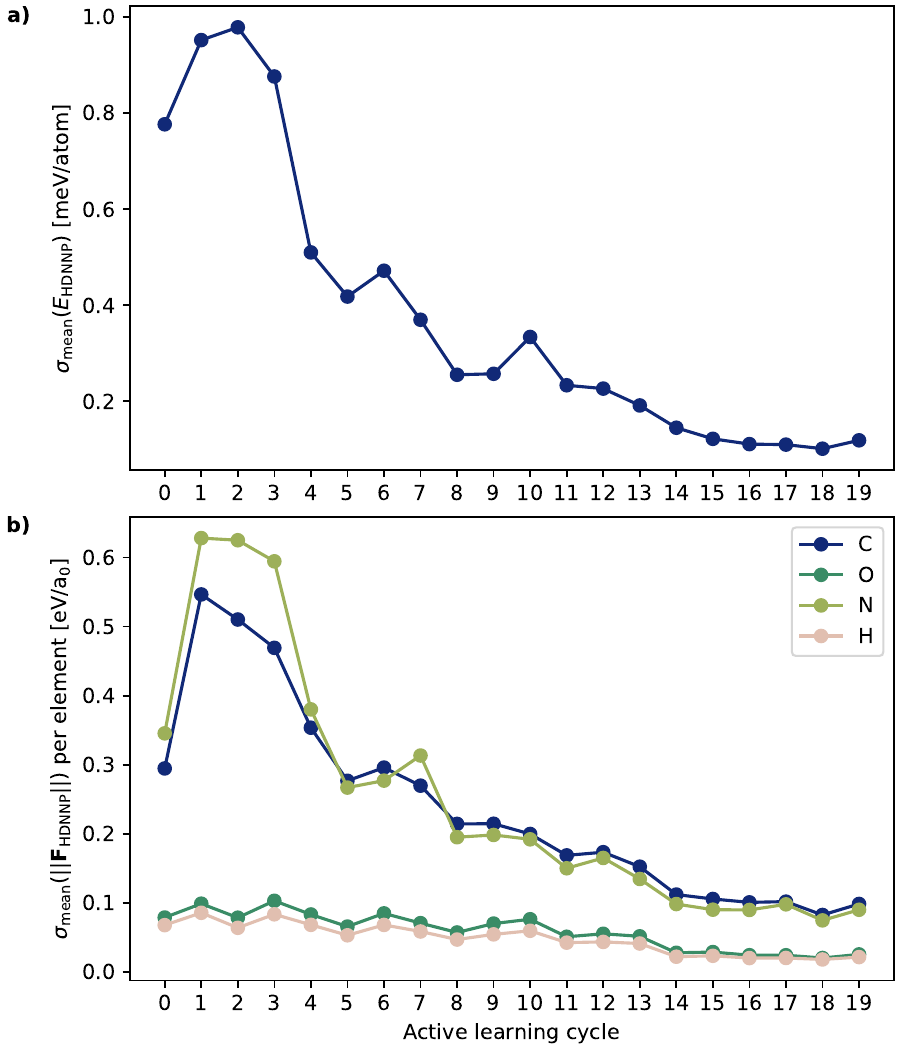}
    \caption{Prediction uncertainties of the energies $E_\mathrm{HDNNP}$ (a) and element-resolved forces $||\mathbf{F}_\mathrm{HDNNP}||$ of a set of 55055 structures along the reaction path during AL. 
    The mean standard deviations $\sigma_{\mathrm{mean}}$ have been computed by averaging the standard deviations $\sigma$ of the predictions of a committee of three HDNNPs over all 55055 structures.
    }
    \label{fig:committee}
\end{figure}

Since the number of structures that can be computed by DFT for validation purposes only (cf. Fig.~\ref{fig:learning_curve}) is limited, we have further tested the reliability of the HDNNPs in each AL cycle for a large number of structures generated from HDNNP-based umbrella sampling simulations. Specifically, we have selected a HDNNP of AL cycle 17 to run 100~ps MD simulations in each of the 55 umbrella sampling windows. We have then chosen structures every 200 time steps to obtain a trial set of 55055 uncorrelated configurations.
Since for these structures no DFT reference values are available, we use the uncertainty of a committee of HDNNPs to estimate the prediction uncertainty. In previous works it has been found\;\cite{kahleQualityUncertaintyEstimates2022, tanSinglemodelUncertaintyQuantification2023, zhuFastUncertaintyEstimates2023, zaverkinUncertaintybiasedMolecularDynamics2024, stolte_random_2024} that prediction errors and uncertainties of HDNNP committees may not be very strongly correlated. Still, high standard deviations of HDNNP committee predictions indicate a too high flexibility of the potentials, because the structures are too different from the geometries included in the training set. The HDNNP committees we employ consist of three members differing in seed and architecture, which have been chosen out of the six HDNNPs fitted in each cycle of AL (cf. Sec.~\ref{subsec: HDNNP_construction}).

The evolution of the committee uncertainty during AL is shown in Fig.~\ref{fig:committee}. It is defined as the standard deviation of the predictions of the committee of HDNNPs averaged over all structures in the trial data set. For the force components this averaging has been carried out by element to obtain information about the element-specific prediction uncertainty (Fig.~\ref{fig:committee}b).
It is clearly visible that the committee uncertainties of the predicted energies and forces decrease and finally converge in the process of AL. While at the beginning the uncertainties first increase due to the inclusion of new high energy structures in less well sampled regions as discussed above, they rapidly decrease again and converge with only marginal fluctuations after cycle 15. Interestingly, for all AL cycles, the energy uncertainty is below 1~meV/atom and thus in the same range as the energy RMSEs (cf. Fig.~\ref{fig:learning_curve}a).

The force prediction uncertainties of the carbon and nitrogen atoms are higher than those of the oxygen and hydrogen atoms. Since the number of water molecules, i.e, the number of oxygen and hydrogen atoms, in the system is much larger than the number of atoms in the reactant molecules, each structure statistically contains more force components of oxygen and hydrogen atoms that are available to optimize the NN weight parameters. It is important to note here that in general atomic forces do not only depend on the weights of the atomic NNs of the atom experiencing the force. Instead, each force component depends  on the atomic NNs of all atoms within the cutoff radius irrespective of the chemical species~\cite{behlerFirstPrinciplesNeural2017}. Due to the large number of solvent molecules, 
the hydrogen and oxygen atomic NNs are predominantly optimized to minimize the force errors of the atoms in the water molecules. Still, they are also very important for the forces acting on the carbon and nitrogen atoms, which have a smaller impact and are thus less accurately described. 
Although this imbalance is not completely removed in the course of AL (cf. Fig.~\ref{fig:committee}b), the uncertainties of the forces of all elements decrease substantially during AL. The decrease of the force uncertainties is much stronger for the nitrogen and carbon atoms since the adaptive Kalman filter optimizer focuses on the optimization of forces with large errors such that carbon and nitrogen forces have an above average impact on the training process. 

Another reason for higher force prediction uncertainties in case of the carbon and nitrogen atoms compared to hydrogen an oxygen atoms could in principle also be slightly higher forces acting on nitrogen and carbon atoms, which could then lead to higher standard deviations. However, normalizing the force standard deviations of each element by the magnitude of the respective forces does not yield more balanced standard deviations of all elements (cf. Fig~S3 in the SI).
 
The results of Fig.~\ref{fig:committee} further show that force uncertainties averaged over all atoms in a system need to interpreted with care since such averaged quantities might be dominated by majority species like the atoms of solvent molecules, while larger errors in a few atoms that are important for the reaction might be overlooked. Here, it is interesting to observe that the uncertainty in the energy prediction seems to be more sensitive to the overall reliability of the potential. 
A comparison of the learning curves in Fig.~\ref{fig:learning_curve} and the uncertainties in Fig.~\ref{fig:committee} shows qualitatively the same systematic improvement of the HDNNPs. However, in the learning curves an abrupt decrease of the RMSEs can be observed in AL cycle 14, i.e., when the cutoff is increased and the ACSFs are augmented by a second set of angular functions. This sudden decrease is not so clearly visible in the uncertainties, since these do not measure the absolute accuracy with respect to DFT but the average standard deviation in predictions of unknown structures. These do not make use of fixed reference values but are more sensitive to the density of training structures, which continuously increases during AL.

\subsection{Stability of MD trajectories} 
\label{subsec:MDval}

\begin{figure}
    \centering
    \includegraphics[width=\columnwidth]{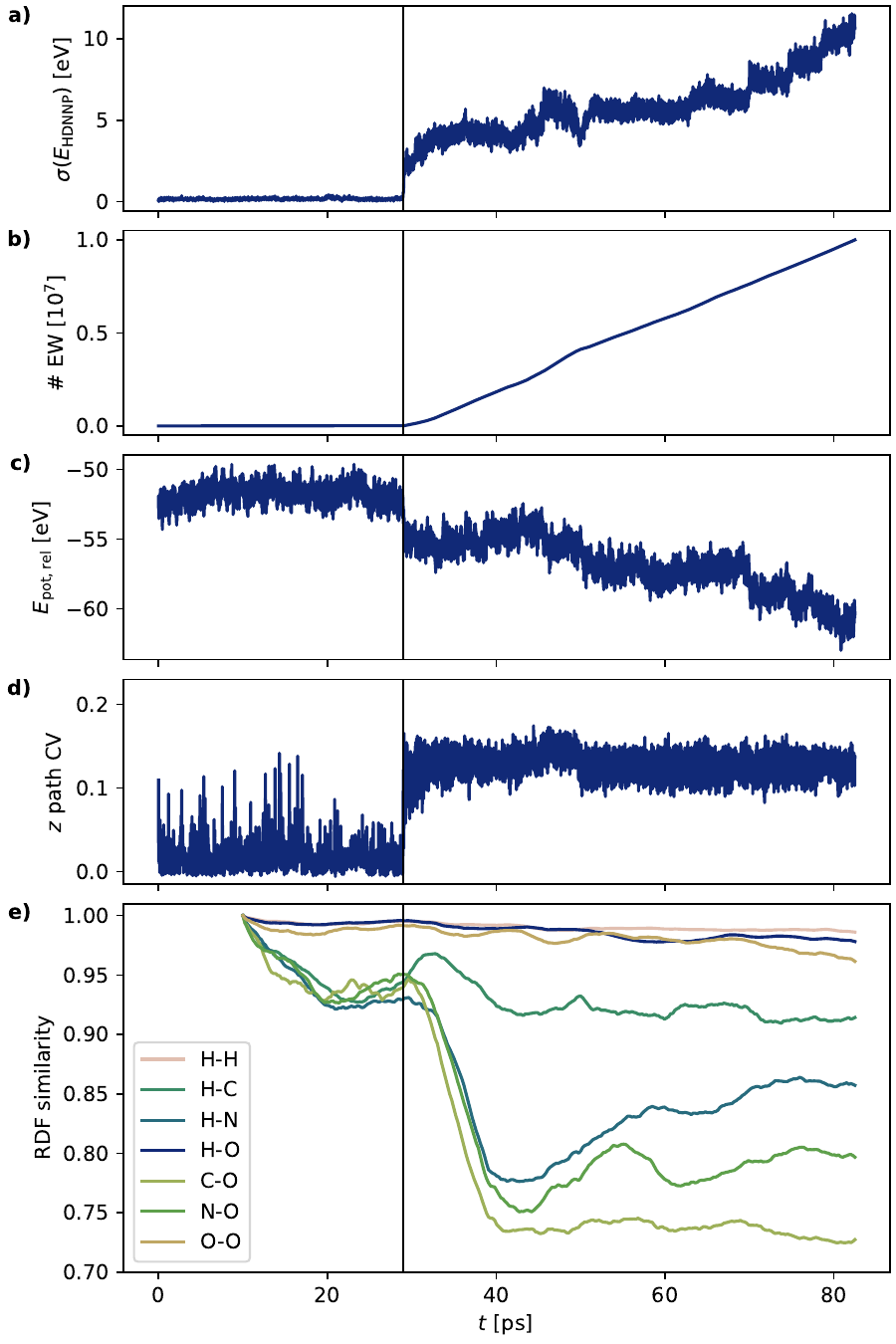}
    \caption{
    Criteria used to assess the stability of MD simulations illustrated for a trajectory at $s = 0.028$ and 300~K generated using a HDNNP of cycle 8. Panel (a) shows the standard deviation $\sigma$ for the energy predicted by a committee of three HDNNPs. In (b) the accumulated number of extrapolation warnings (EW) for atomic environments exhibiting ACSF values beyond the training range is given. The sudden drop in the potential energy in (c) results from a structural change that is also visible in the path CV $z$ in panel (d). This structural change is also the reason for the approximately constant number of EWs emerging in each MD step afterwards. Structural changes can also be identified in the radial distribution function similarity scores (Eq. \ref{eq:score_eq}) between different fragments of the trajectory in panel (e).
    All these criteria consistently lead to approximately the same time at about 29~ps when the simulation becomes unreliable (vertical line).
     }
    \label{fig:stable_simulation_length}
\end{figure}

After examining the construction of the reference dataset and the prediction accuracy of the HDNNPs we now assess the reliability of the potentials in MD simulations and show how AL improves the stability of the obtained trajectories. This test represents an important validation step since in particular the performance for structures not encountered in the training process is crucial for the applicability of a potential. Inaccurate forces might guide the system away from the explored regions of configuration space, giving rise to unphysical geometries and thus wrong trajectories. 

Apart from analyzing the uncertainty of predictions as discussed above, unseen parts of configuration space can be identified by monitoring the extrapolation of the ACSFs describing the atomic environments. Extrapolations are present if symmetry function values outside the range of values spanned by the configurations of the training set emerge. Such extrapolations do not necessarily lead to unphysical trajectories, but if a large number of extrapolations is observed, trajectories should be terminated and the potential should be extended by adding more data, either to include important but missing parts of configuration space, or to close ``holes'' in the PES which would allow the system to enter unphysical regions.

We will now explore, how unphysical trajectories can be recognized and which criteria next to ACSF extrapolations can be used for this purpose. In particular, we will focus on identifying the simulation time step, after which trajectories become unreliable, and how the length of a reliable trajectory extends with AL. For this purpose, we monitor the evolution of a variety properties during the simulation. These properties are the uncertainty of the energy prediction of a committee of HDNNPs, the accumulated number of ACSF extrapolations, the potential energy of the system, the $z$ path CV, and a similarity score to compare radial distribution functions (RDFs) $g$ at different time steps.

The RDF similarity score up to a distance $L$ can be defined based on a comparison of the RDF for a trial trajectory and an equilibrium (EQ) reference RDF~\cite{schranConvergedColoredNoise2018,schranMachineLearningPotentials2021a} as
\begin{equation}
\mathrm{score} = 1 - \frac{\int_0^{L} \left| g^{\mathrm{EQ}}(r) - g^{\mathrm{trial}}(r) \right| \mathrm{d}r}{\int_0^{L} \left|g^{\mathrm{EQ}}(r)\right| \mathrm{d}r + \int_0^{L} \left|g^{\mathrm{trial}}(r) \right|\mathrm{d}r},
\label{eq:score_eq}
\end{equation} 
which yields a score of one if the two RDFs are fully identical and lower values otherwise. 
Since in the present case we are interested in changes of the RDF with progressing simulation time, we compare the RDF of the first $10$\;ps of the simulation, $g^{\mathrm{EQ}}$, as reference and the $10\;$ps before the respective simulation timestep of interest, $g^{\mathrm{trial}}$.

Figure\;\ref{fig:stable_simulation_length} shows all investigated properties as a function of the simulation time for the example of a trajectory generated with a HDNNP of AL cycle 8 along the reaction path at $s$=0.028 and 300~K. Panel (a) shows the committee uncertainty of the predicted total energy, which remains very low until a simulation time of about 29~ps has been reached. Then, the uncertainty strongly increases indicating the presence of atomic configurations that strongly deviate from the underlying training set. At the same time also the number of accumulated extrapolations shown in panel (b) starts to increase almost linearly, indicating that in each of the geometries visited in the remaining part of the trajectory about the same number of about 100 extrapolations per step occurs. In panel (c) the relative potential energy of the system is plotted, which in agreement with the increased uncertainty shows a sudden decrease at 29~ps followed by relatively large variations with time. The related structural change is also visible in the path CV variable $z$ in panel (d) indicating a structural deviation of the system from the reaction path.
The evolution of the RDF similarity scores in panel (e), which are computed starting from simulation time 10~ps onwards due to the required minimum sampling time, first decrease slowly and then equilibrate around a value of 0.95 at a simulation time of 20~ps. This deviation from one is expected since the rather limited sampling time of only 10~ps introduces random noise in the RDFs that fluctuates with simulation time. The solvent scores of the O-O, H-O and H-H RDFs stay closer to values of one since they are based on much better statistical sampling. Then, the scores of most RDFs exhibit strong changes occurring at about 29~ps. Again, only the O-O, H-O and H-H RDF scores, which are dominated by the solvent, do not show this drop strongly indicating that the structural change in the system is primarily related to the less well-represented reactants. 

Overall, all properties investigated in Fig.~\ref{fig:stable_simulation_length} consistently indicate a sudden structural rearrangement in the system, which sets in after approximately 29~ps and seems to be irreversible. After such a transition in the system the trajectory can be defined as ``unstable'' and cannot be used in production MD simulations. Apart from this trajectory representing an example in the reactant basin at $s$=0.028, the detailed analysis of two further example trajectories close to the transition state and in the product basin can be found in Fig.\;S44 and Fig.\;S45 in the SI.
Due to the similar information content of all indicators in Fig.~\ref{fig:stable_simulation_length}, from now on we will use an accumulated number of 10,000 extrapolations as indicator for unstable trajectories that will then be stopped and discarded, a criterion that is readily available in the n2p2 code~\cite{LibraryBasedLAMMPSImplementation}. This number of extrapolations allows to continue trajectories after a few extrapolations per MD step, which are unavoidable in the course of a simulation and usually do not result in wrong trajectories.\\

\begin{figure}
    \centering
    \includegraphics[trim={2cm 0cm 1cm 2cm},clip,width=\columnwidth]{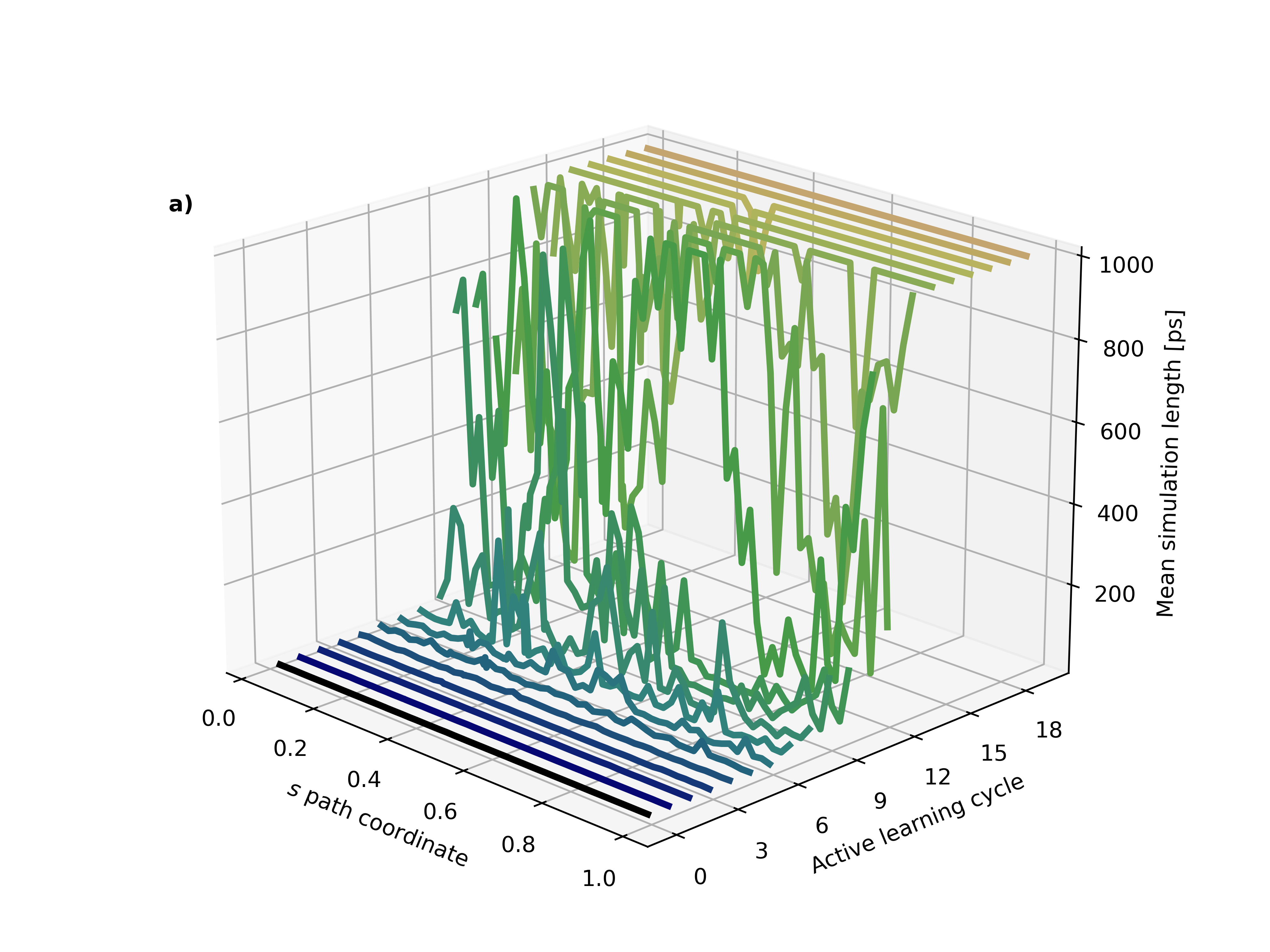}
    \includegraphics[width=\columnwidth]{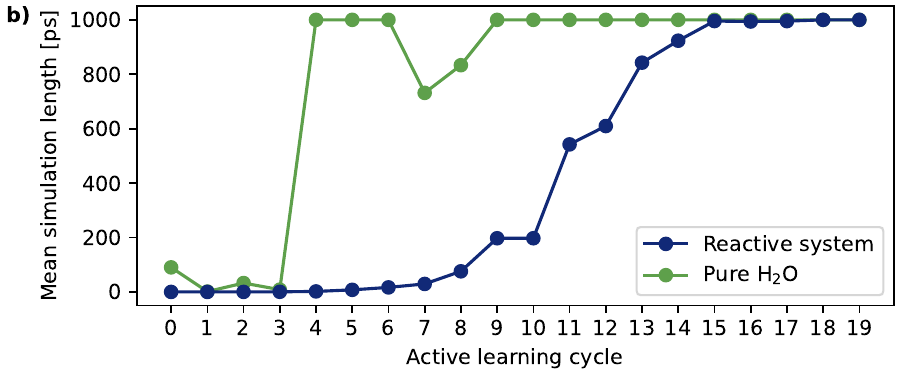}
    \caption{Mean ``stable'' simulation times for HDNNP-driven MD trajectories. Panel (a) shows the simulation times in all 55 umbrella sampling windows along the path CV $s$ averaged over four different trajectories as a function of the AL cycle (220 trajectories in total). In panel (b) the average simulation times for pure water (averaged over four trajectories) and for the reactive system (averaged over the 220 trajectories of all windows) are compared showing that the sampling of the pure solvent is essentially completed after four AL cycles.
    The simulations have been defined as ``unstable'' using a criterion of 10,000 accumulated extrapolations in the ACSF space. Otherwise, they have been considered stable and terminated when a simulation time of 1\,ns has been reached.}
    \label{fig:length}
\end{figure}

Having defined a criterion of the stability of a trajectory, we can now compare how the stable simulation length evolves during the AL process. Figure\;\ref{fig:length}a shows the average stable simulation length for the reactive system averaged over four independent trajectories employing two different HDNNPs and two different random number seeds for the velocity initialization. The data is given for each umbrella sampling window as a function of the AL cycle. In each AL cycle the same 55 structures have been used to start the simulations. If no instability has been observed up to a runtime of 1~ns, the simulations have been terminated and considered as long-term stable.

After an initially very slow increase in stable simulation time, starting in AL cycle seven the MD simulations of the reactive system become longer in each cycle and converge approximately in cycle 15 reaching the defined maximum simulation length of 1~ns for most windows. This finding is in good agreement with our earlier analysis showing that the relevant configuration space is essentially covered at this stage of AL. 
A two-dimensional projection of the runtimes for all umbrella sampling windows in all AL cycles is shown in Fig.~S46 in the SI. From this plot and also from Fig.~\ref{fig:length} it can be concluded that some windows along the reaction path need more AL cycles to reach stable trajectories than others. These regions can be found, e.g., around $s=0.2$ and at $s=0.75$. 

Figure\;\ref{fig:length}b shows the overall stable simulation length of the reactive system averaged over the in total 220 trajectories of all 55 umbrella sampling windows shown in Figure\;\ref{fig:length}a and for comparison the stable simulation time of pure water in a periodic box containing 160 molecules. For pure water, for each AL cycle the lengths of four simulations employing two different HDNNPs and two different seeds for the velocity initialization have been averaged. Although the dataset does not contain any pure water structures, bulk-water like atomic environments are well represented in the dataset such that the obtained HDNNPs can also be used to run MD simulations of this system. It can be seen that the stable simulation times of pure water converge much faster than for the reactive system, i.e., essentially at cycle four of the AL process. This clearly demonstrates that the water degrees of freedom are already well sampled in the early AL cycles containing only MD simulations of the initial and final window. 

In summary, the evolution of simulation times suggests full convergence of AL for the reactive system after 18 cycles when all simulations reach 1\;ns. Still, the long-term stability of the trajectories according to our definition given above alone does not provide evidence that all trajectories are physically reliable. For instance, small numbers of extrapolations might still occur. Such encounters are labeled as extrapolation warnings and during MD simulations the corresponding atom, the particular ACSF and the magnitude of the extrapolation can be stored and analyzed.\\

\begin{figure}
    \centering
    \includegraphics[width=\columnwidth]{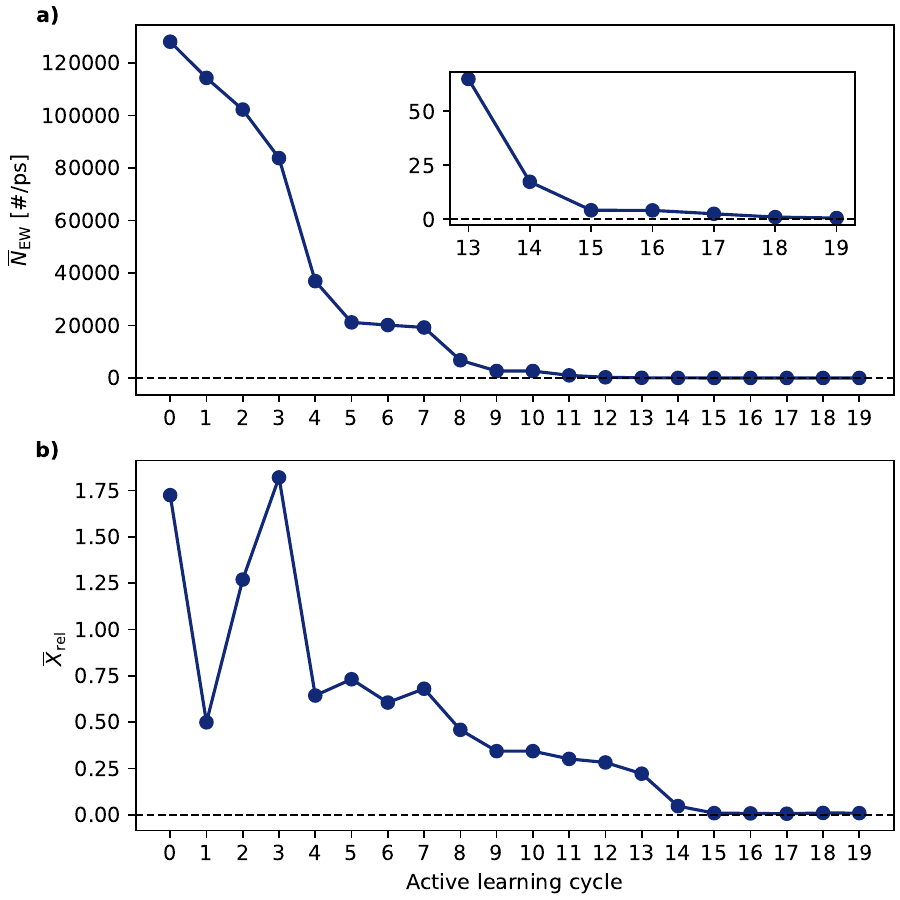}
    \caption{Number of extrapolations in MD simulations as a function of the AL cycle. Panel (a) shows the average number of extrapolations per picosecond simulation time. Panel (b) compiles the average relative magnitude of the extrapolations $\overline{X}_{\mathrm{rel}}$ calculated using Eq.\;\ref{eq:EW_mag}. For each AL cycle, the numbers and magnitudes of extrapolations have been averaged over the same 220 trajectories analyzed in Fig.~\ref{fig:length}.}
    \label{fig:extrapolation_warnings}
\end{figure}

In Fig.\;\ref{fig:extrapolation_warnings} the average number of extrapolations per picosecond simulation time and the relative magnitudes of the observed extrapolations are shown averaged for all umbrella sampling simulation along the reaction pathways for all AL cycles. The number of extrapolations $N_\mathrm{EW}$ is counted for each trajectory, divided by the simulation length and averaged for all simulations of the respective AL cycle. It can be seen that the resulting average number of extrapolation warnings in Fig.~\ref{fig:extrapolation_warnings}a strongly decreases as the AL progresses. It plateaus from cycle five to seven, but then continues to decrease and converges to very low values in cycle 15.

As the absolute magnitude of the extrapolations strongly depends on the range of the respective symmetry function values, the extrapolating value $G_i^\mathrm{X}$ is normalized by the range spanned by the largest symmetry function value $G_i^\mathrm{max}$ and the smallest symmetry function value $G_i^\mathrm{min}$ of the respective function in the reference dataset as 

\begin{align}
    X_\mathrm{rel} =\left\{\begin{array}{cll}
    \frac{G^\mathrm{X}_i-G_i^\mathrm{max}}{G^\mathrm{max}_i - G_i^\mathrm{min}} & \text { for } & G^X_i >  G_i^\mathrm{max}\\
    &&\\
    \frac{G_i^\mathrm{min} - G_i^\mathrm{X}}{G_i^\mathrm{max} - G_i^\mathrm{min}} & \text { for } & G^X_i < G_i^\mathrm{min}\,.
    \end{array}\right.
    \label{eq:EW_mag}
\end{align}

The magnitudes of all extrapolations are then calculated and averaged for all MD simulations resulting the average magnitudes of extrapolations $\overline{X}_\mathrm{rel}$. As shown in Fig.~\ref{fig:extrapolation_warnings}b, at the beginning of the AL process the average magnitudes fluctuate strongly, then decrease continuously after cycle four and converges to essentially zero after cycle 15. The decrease of the average extrapolation magnitude has two reasons: First, with the exploration of configuration space and adding more data during the AL, the range of the ACSFs extends and hence less extrapolations are observed. Second, with improving HDNNPs the number of unreliable force predictions, which could drive the system to configurations which are usually not visited in MD simulations, strongly decreases. Notably, in the beginning of the AL process the average magnitude of extrapolations decreases strongly due to an increased range of ACSF after the first AL cycle. After this cycle, however, the extrapolation magnitude increases again for two cycles. At this point the MD simulations run longer and thereby allow to explore a larger configuration space increasing the probability of larger extrapolations.
Afterwards, starting in cycle four the magnitude and number of extrapolations decrease as the dataset increasingly converges and the HDNNPs become more reliable.\\

\begin{figure}
    \centering
    \includegraphics[width=\columnwidth]{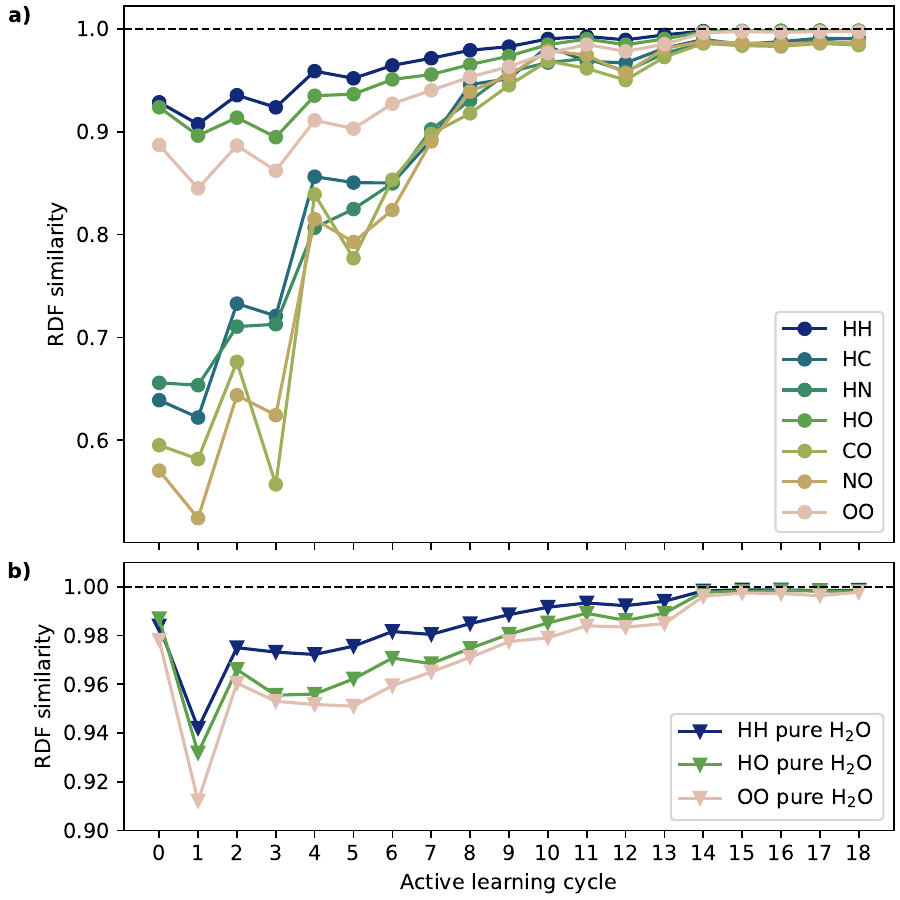}
    \caption{Convergence of the radial distribution functions (RDF) of the reactive system (a) and of pure water (b) as a function of the AL cycle. The convergence is measured by the RDF similarity score given in Eq.~\ref{eq:score_eq} using the RDF of AL cycle 19 as reference.
    }
    \label{fig:RDF_sim}
\end{figure}

Apart from monitoring the number of extrapolations, for validating the quality of the HDNNPs it is necessary to directly assess the convergence of physical properties of the system with RDFs being important examples to characterize the structure of the system. Figure\;\ref{fig:RDF_sim} shows the RDF similarity scores as a function of the AL cycle for the reactive system in panel (a) and for pure water in panel (b). The similarity scores (Eq.~\ref{eq:score_eq}) are computed with respect to converged RDFs obtained with the final potential of cycle 19. In particular pure water is an interesting test case since RDFs computed with the same exchange-correlation function have been reported in the literature\;\cite{forster-tonigoldDispersionCorrectedRPBE2014,P4556} and are in excellent agreement with our work. As the RDF similarity score in Fig.\;\ref{fig:RDF_sim}b demonstrates, selecting a larger cutoff is important to obtain a converged RDF of liquid water.

As the RDF of the reacting system is expected to change along the reaction path, the MD simulations for computing the RDF were run at the reactant basin at $s=0.9$. The RDFs for CC, CN and NN have not been computed as the low numbers of atomic pairs of these element combinations do not allow to obtain statistically meaningful RDFs. The RDF similarity scores in Fig.\;\ref{fig:RDF_sim}a show good convergence and reach their final values with the increase of the cutoff to 10 a$_0$ in cycle 14. As observed before, a more rapid convergence is found for the HH, HO and OO RDFs due to the better description of water in the early AL cycles. This difference between the similarity score of the HH, HO and OO RDFs and the RDFs centered on carbon and nitrogen is higher at the beginning of the AL process and decreases as the AL progress continues. Moreover, it should be noted that MD simulations in early AL cycles are less stable, and the resulting shorter simulation times introduce some noise in the RDFs preventing high similarity scores in the first AL cycles.

\subsection{Free energy profile} \label{subsec:free_energy}

\begin{figure}
    \centering
    \includegraphics[width=\columnwidth]{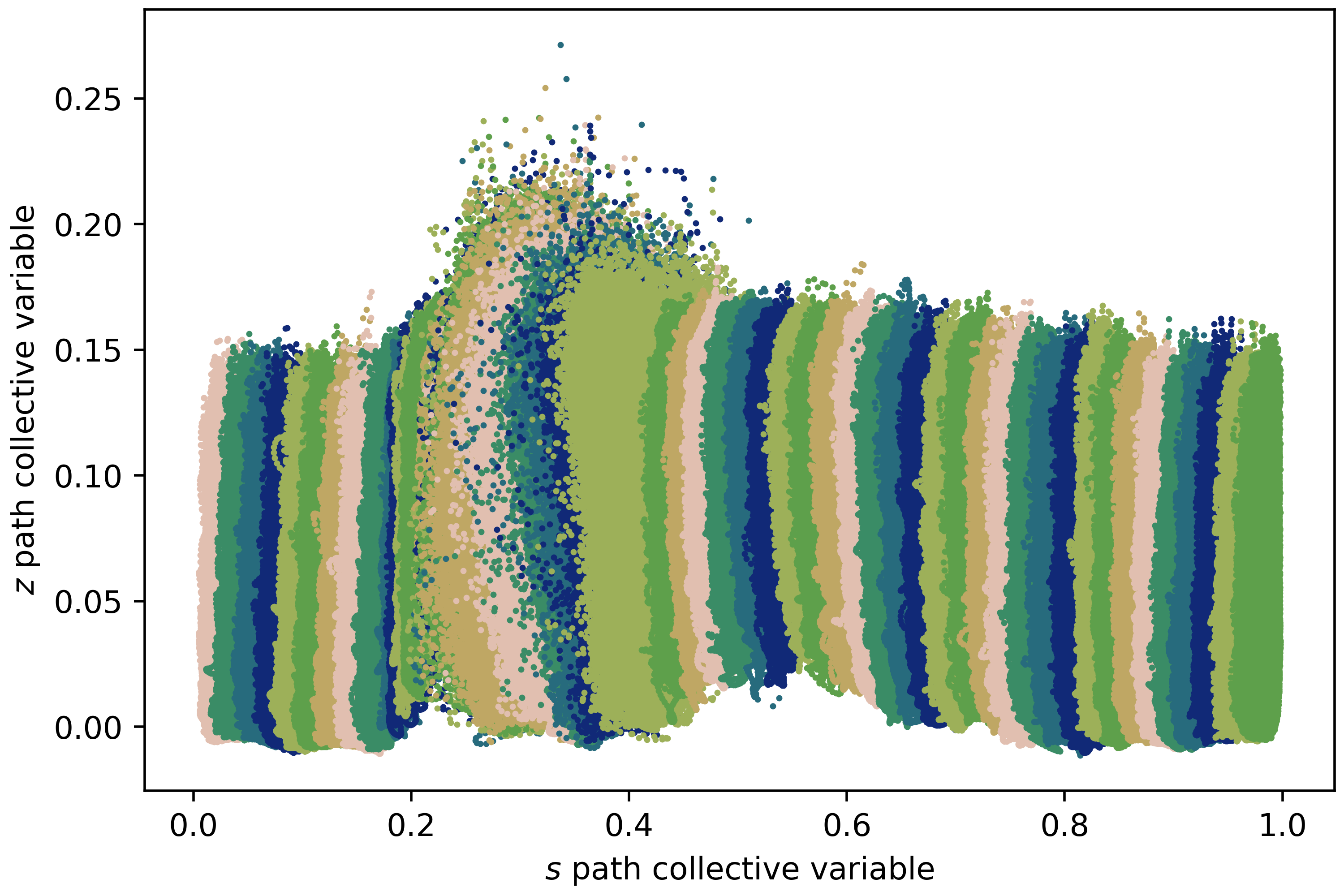}
    \caption{Distribution of configurations in the ($s,z$) path CV space visited in the umbrella sampling simulations employing a HDNNP of AL cycle 19. The points have been colored to distinguish subsequent umbrella sampling windows. Disconnected areas of the same color correspond to different windows. 
    }
    \label{fig:s_z_MD}
\end{figure}

\begin{figure}
    \centering
    \includegraphics[width=\columnwidth]{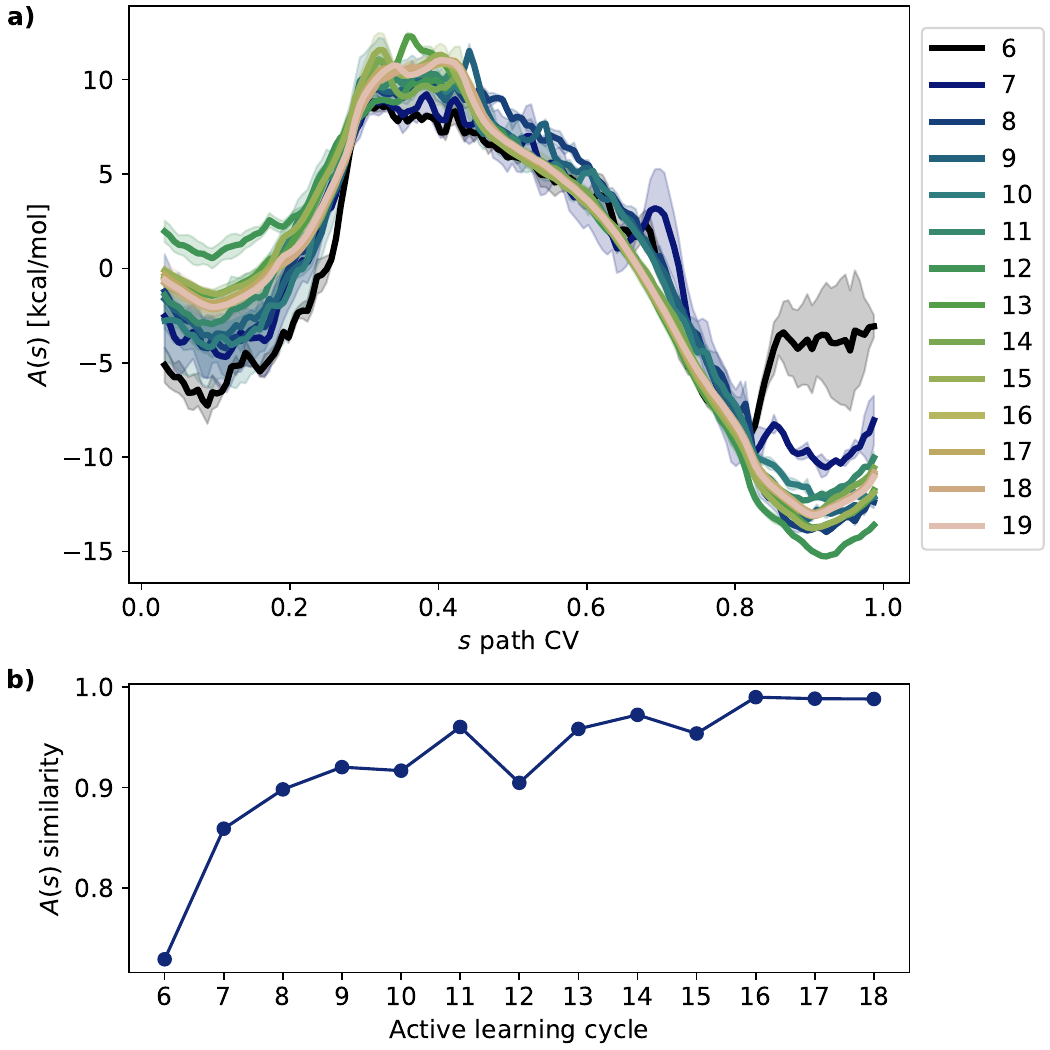}
    \caption{
    Convergence of the free energy $A$ obtained with umbrella sampling simulations along the path CV $s$ for the different AL cycles (a). In Panel (b) the convergence of the free energy profiles with respect to the free energy profile of cycle 19 is measured with the similarity score of Eq.\;\ref{eq:score_eq}. Free energy profiles for cycles 0 to 5 are not shown because the runtimes of the MD simulations are too short (cf. Fig.~\ref{fig:length}).  
    The statistical uncertainty of the free energy is given for each cycle by the transparent areas of the respective color. 
    }
    \label{fig:free_energy}
\end{figure}

Finally, we determine the free energy profiles of the first reaction step of the Strecker synthesis of glycine by umbrella sampling simulations using HDNNPs obtained in different AL cycles. The free energy profiles are computed based on trajectories of 1\;ns length. For AL cycles earlier than cycle 16, stable simulations times are typically shorter (cf.~Fig.~\ref{fig:length}) and in these cases the shortest stable simulation time of the umbrella sampling windows was applied to all windows for consistency. The trajectories were evaluated to estimate the uncertainty of the free energy profile as explained in section~\ref{subsec:MD}.

The distribution of configurations visited in the umbrella sampling simulation in the ($s,z$) path CV space of cycle 19 in Fig.\;\ref{fig:s_z_MD} shows a dense overlap of the sampling windows and a good spatial confinement of the individual MD simulations in the respective windows.

The resulting free energy profiles are shown in Fig.\;\ref{fig:free_energy}a for cycles 6 to 19. The simulation lengths of earlier cycles were found insufficient to calculate the corresponding free energy profiles. 
It can be observed that the convergence of the free energy towards the end of the AL process is excellent for the whole reaction with only minor deviations of 0.5~kcal/mol at the transition state.
As reported in previous work, the reason for this uncertainty is a small hysteresis~\;\cite{magrino_step_2021} caused by the path CVs, but it is not related to the accuracy of the HDNNPs employed in the present work. Overall, the free energy profile is very well converged, which is also confirmed by the similarity scores shown in Fig.~\ref{fig:free_energy}b. 

The free energy activation barrier between reactants and transition state as well as the free energy differences between reactants and products are compiled in Table\;\ref{TAB:DA} and compared to previous \emph{ab initio} MD work. In spite of the different exchange-correlation functional employed in the DFT calculations of Ref.~\;[\citenum{magrino_step_2021}] (PBE functional~\cite{perdewGeneralizedGradientApproximation1996a} with D2 van der Waals corrections\;\cite{grimmeSemiempiricalGGAtypeDensity2006}), overall there is very good agreement in the obtained free energies.

\begin{center}
\begin{table}
 \caption{Activation barriers $\Delta A^\dagger$ and free energy differences $\Delta A$ between reactants and products obtained in this study and the \textit{ab initio} MD study of Ref.\;[\citenum{magrino_step_2021, huetCorrectionStepStep2024}]. 
 }
\label{tab_eta_rad_mn3_2}
\vspace{0.25cm}
\begin{tabular}{ccc}
\toprule
& This work & Literature\\
\midrule
$\Delta A^\dagger$ & 12.2 & 14 (Ref.\;[\citenum{magrino_step_2021}])\\
$\Delta A$ & -11.1 & -11.2 (Ref.~[\citenum{huetCorrectionStepStep2024}])\\
\bottomrule
\end{tabular}
\label{TAB:DA}
\end{table}
\end{center}
%

\section{Conclusions}

In this work a systematic protocol has been presented to construct and validate a HDNNP for studying the first step of the Strecker synthesis of glycine as a prototypical case for a chemical reaction in an explicit solvent. The potential allows to accurately determine a converged free energy profile with DFT accuracy by performing extended umbrella sampling simulations along the reaction path. Central to our approach is an iterative active learning process to determine new data points for the training set of the potential. While active learning is commonly used in the construction of MLPs, here we go beyond the typical analysis restricted to energy and force errors and present detailed insights into this process by rigorously monitoring the increasing stability and accuracy of the HDNNP and of the obtained results using different physical properties. 
 
Starting with equilibrium MD simulations in the reactant and product basins we demonstrate that an accurate description of the pure solvent can be obtained in an early stage of active learning. 
However, the accurate representation of the reaction path requires sampling new reference structures in a series of systematic simulations in umbrella sampling windows, which has been rarely done to date in combination with active learning.

Dimensionality reduction techniques can be employed to visualize the progress in mapping the diversity of atomic environments in particular for the reactant molecules. The accuracy of the HDNNP for the structures along the reaction path can be monitored by investigating energy and force RMSEs with respect to the available reference DFT data, or using the prediction uncertainty of a committee of HDNNPs for a much larger pool of validation structures generated by HDNNP-driven simulations. The accuracy in the representation of the atomic forces depends on the relative abundance of the respective element in the systems, which leads to a better description of the majority of solvent molecules in the early phase of active learning. To a large extent, the quality of the forces can be balanced by the optimization algorithm of the neural network and by converging the dataset size. Moreover, it was found that in particular at the beginning of active learning new configurations at the boundaries of the explored configuration space are sampled, while at later stages the remaining gaps are filled as the potential and dataset converge. Finally, a sufficiently large cutoff in combination with a reasonable set of atom-centered  symmetry functions is required to obtain accurate potentials, while a less stringent description of the atomic environments can speed-up the initial phase of active learning, which allows to determine a close-to converged dataset with reduced computational costs.

The dominance of solvent molecules in the system creates a compositional imbalance, which restricts the usefulness of averaged metrics such as RMSE values and prediction uncertainties in assessing the quality of the potential. Therefore, another important target is the long-term stability of MD trajectories, which we have investigated using a variety of criteria that turned out to provide a very consistent measure for the quality of a trajectory. An easy to apply criterion is the number of extrapolations beyond the range of atom-centered  symmetry function values describing the atomic environments in the training set.
However, since even the long-term stability of trajectories is insufficient to ensure a correct physical description of the system, finally, we have assessed the convergence of physical properties like radial distribution functions and the free energy profile with increasing dataset size.

Overall, we find that a hierarchical approach consisting of the assessment of errors and uncertainties, the long-term stability of trajectories and monitoring physical properties allows to construct high-quality HDNNPs suitable for studying molecular systems in solution.

\section*{Supplementary Material}

The supplementary material contains detailed information of settings used for the construction of the HDNNPs and additional information and insights of the active learning process.
\\
\\
\begin{acknowledgments}
A.M.T. and J.B. are grateful for support by the Deutsche Forschungsgemeinschaft (DFG) (BE3264/16-1, project number 495842446 in priority program SPP 2363 ``Utilization and Development of Machine Learning for Molecular Applications – Molecular Machine Learning'') and under Germany's Excellence Strategy – EXC 2033 RESOLV (project-ID 390677874). T.D. is grateful for the funding of the European Union - NextGenerationEU initiative and the Italian National Recovery and Resilience Plan (PNRR) from the Ministry of University and Research (MUR), under Project PE0000013 CUP J53C22003010006 "Future Artificial Intelligence Research (FAIR)".
A.M.S. is grateful to French supercomputing facilities GENCI for grants A140901387 and A160901387.
\end{acknowledgments}

\section*{Author Declaration}
\subsection*{Conflict of interests}
The authors have no conflicts to disclose.

\bibliography{literature}

\end{document}